\documentclass{emulateapj}

\input psfig.sty

\lefthead{Alonso-Herrero et al.}
\righthead{Local LIRGs}
\begin{document}



\title{Near-Infrared and Star-forming properties of Local Luminous Infrared
  Galaxies}

\author{Almudena Alonso-Herrero\altaffilmark{1},
George H. Rieke\altaffilmark{2},
Marcia J. Rieke\altaffilmark{2}, 
Luis Colina\altaffilmark{1}, 
Pablo G. P\'erez-Gonz\'alez\altaffilmark{2},
and Stuart D. Ryder\altaffilmark{3}}
\altaffiltext{1}{Departamento de Astrof\'{\i}sica Molecular e Infrarroja,
  Instituto de Estructura de la Materia, CSIC, Serrano 121, 
28006 Madrid, Spain; e-mail: aalonso@damir.iem.csic.es}
\altaffiltext{2}{Steward Observatory, The University of Arizona, 
933 N. Cherry Avenue, Tucson, AZ 85721}
\altaffiltext{3}{Anglo-Australian Observatory, PO Box 296, Epping, NSW 1710, Australia}

\begin{abstract}
We use {\it Hubble Space Telescope} ({\it HST}) 
NICMOS continuum and Pa$\alpha$ observations  to study 
the near-infrared and star-formation properties of a representative
sample of 30 local ($d\sim 35-75\,$Mpc)  luminous infrared galaxies
(LIRGs, infrared [$8-1000\,\mu$m] luminosities of $\log L_{\rm
  IR}=11-11.9\,[{\rm L}_\odot]$). The data provide spatial
resolutions of $25-50$\,pc and cover the central $\sim 3.3-7.1\,$kpc regions
of these galaxies.  About half of the LIRGs show compact ($\sim 1-2\,$kpc) 
Pa$\alpha$ emission with a high surface brightness in the form of 
nuclear emission, rings, and mini-spirals. The rest of the sample show
Pa$\alpha$ emission along the disk and the spiral arms extending over scales
of $3-7\,$kpc and larger. About half of the sample contains HII regions with
H$\alpha$ luminosities significantly higher than those observed in normal
galaxies.  There is a linear
empirical relationship between the mid-IR $24\,\mu$m and hydrogen recombination
(extinction-corrected Pa$\alpha$) luminosity for these LIRGs, and the HII
regions in the central part of M51. This relation holds over more than four decades
in luminosity suggesting that the mid-IR emission is a good tracer of the 
star formation rate (SFR). Analogous to the widely used relation between the SFR  and total
IR luminosity of Kennicutt (1998), we derive an empirical calibration of the 
SFR in terms of the monochromatic $24\,\mu$m luminosity that can be used 
for luminous, dusty galaxies.

\end{abstract}


\section{Introduction}

The importance of infrared (IR) bright galaxies 
has been recognized since their discovery more than 30 years ago
(Rieke  \& Low 1972) and the detection of large numbers by the {\it IRAS} satellite (Soifer et
al. 1987). Specifically, there has been controversy over
the extent to which luminous and ultraluminous IR galaxies (LIRGs, 
$L_{\rm IR}=L[8-1000\mu {\rm m}] = 10^{11}-10^{12}\,{\rm L}_\odot$, 
and ULIRGs $L_{\rm IR} > 10^{12}\,{\rm L}_\odot$, respectively) in the local universe
are powered by intense star
formation as opposed to active galactic nuclei (see the review by Sanders \& Mirabel 1996).
The process of merging with accompanying super-starbursts that produces LIRGs and ULIRGs 
appears to be an
important stage in galaxy evolution, possibly even converting spiral
galaxies into ellipticals (e.g., Genzel et al. 2001;  Colina et al. 2001, 
and references therein) and giving rise to quasars (Sanders et al. 1988).

In recent years with the advent of the new generation of 
IR satellites, there has been a
considerable effort in understanding the properties of distant IR-selected
galaxies. The majority of these galaxies at $z<1$ are in the LIRG class,
and they  make a significant contribution 
to the star formation rate density at $0.5<z<2$ (Elbaz et al. 2002; 
P\'erez-Gonz\'alez et al. 2005; Le
Floc'h et al. 2005). 
At $z\sim 0.7 - 1$ a significant fraction of the LIRG population appears to be
morphologically disturbed spiral galaxies (Bell et al. 2005; Shi et al. 2006),
rather than interacting and merging systems. 
These high-$z$ IR bright galaxies show morphologies and properties (see e.g.,
Zheng et al. 2004; Papovich et al. 2005; Melbourne, Koo, \& Le Floc'h 2005; 
Shi et al. 2006) typical of 
low-$z$ LIRGs and ULIRGs 
(among others, Murphy et al. 1996;  Surace, Sanders, \& Evans 2000, 
Scoville et al. 2000;  
Alonso-Herrero et al. 2002; Colina et al. 2001;  Colina, Arribas, \&
Clements 2004; Arribas et al. 2004). Lagache,
Puget, \& Dole (2005) have recently reviewed the properties of
the high-$z$ IR galaxy population.

The local density of LIRGs is two 
orders of magnitude higher than that of ULIRGs (e.g.,
Soifer et al. 1987).  
Moreover, Murphy et al. (2001) have argued that in the local
universe interacting galaxies may spend a significant fraction of
their lifetime as LIRGs, whereas the ULIRG phase may just be short and
recurrent during the merging process.
However, the majority of the detailed studies of the local universe IR galaxies 
have been focused on the objects
with the largest IR luminosities (i.e., ULIRGs).
The LIRGs have been generally neglected, except for detailed studies 
of a few famous examples (e.g., Gehrz, Sramek, \& Weedman 1983; 
Doyon, Joseph, \& Wright 1994; 
Genzel et al. 1995; Satyapal et al. 1999;
Sugai et al. 1999; L\'{\i}pari et al. 2000, 2004; 
Alonso-Herrero et al. 2000, 2001).
Thus, understanding  low-$z$ LIRG samples is critical for  
interpreting the properties of IR-selected high-$z$ galaxy populations; in addition, they 
represent the link between ULIRGs and the population of field galaxies
as a whole.

\begin{deluxetable*}{llcclcccc}

\tiny
\tablewidth{18cm}
\tablecaption{The sample of local universe LIRGs.}


\tablehead{Galaxy Name  &
IRAS Name &
$v_{\rm hel}$ &  
Dist & $\log L_{\rm IR}$ & 
Spect. &
Refs  &
Morphology\\
 & & (km s$^{-1}$) & 
(Mpc) & (L$_\odot$) & class\\
(1) & (2) & (3) &(4) & (5) & (6) & (7) & (8)}
\startdata

NGC~23 & IRASF~00073+2538 & 4566  & 59.6 & 11.05 &H\,{\sc ii}   &1 & paired
with NGC~26\\
MCG~+12-02-001      & IRASF~00506+7248 & 4706  & 64.3 & 11.44  &--    &-- & in group?\\
NGC~633            & IRASF~01341$-$3735 & 5137  & 67.9 & 11.09: &H\,{\sc ii} &2,3 &
paired with ESO~297-G012 \\
UGC~1845           & IRASF~02208+4744 & 4679  & 62.0 & 11.07  &--   &-- & isolated\\
NGC~1614           & IRASF~04315$-$0840 & 4746  & 62.6 & 11.60  &H\,{\sc ii}   &1,2,3 & merger\\
UGC~3351           & IRASF~05414+5840 & 4455  & 60.9 & 11.22  &Sy2   &4,5 &
paired with UGC~3350\\
NGC~2369           & IRASF~07160$-$6215 & 3237  & 44.0 & 11.10  &--    &--  & isolated\\
NGC~2388           & IRASF~07256+3355 & 4134  & 57.8 & 11.23: &H\,{\sc ii}
&1  & paired with NGC~2389\\
MCG~+02-20-003      & IRASF~07329+1149 & 4873  & 67.6 & 11.08: &--    &--  & in group\\
NGC~3110           & IRASF~10015$-$0614 & 5034  & 73.5 & 11.31: &H\,{\sc ii}
&2 & paired with MCG~$-$01-26-013\\
NGC~3256           & IRASF~10257$-$4339 & 2814  & 35.4 & 11.56  &H\,{\sc ii}
& 6 & merger\\
NGC~3690/IC~694     & IRASF~11257+5850 & 3121  & 47.7 & 11.88  &Sy2-H\,{\sc ii}
  & 7 &
close interacting pair\\
ESO~320-G030       & IRASF~11506$-$3851 & 3232  & 37.7 & 11.10  &H\,{\sc ii}   & 8 & isolated\\
MCG~$-$02-33-098E/W   & IRASF~12596$-$1529 & 4773  & 72.5 & 11.11  &H\,{\sc ii} (both)  &1,3 &
close interacting pair\\ 
IC~860             & IRASF~13126+2453 & 3347  & 59.1 & 11.17: &H\,{\sc ii}     &5 & isolated\\
NGC~5135           & IRASF~13229$-$2934 & 4112  & 52.2 & 11.17  &Sy2   &2,3 & in group\\
NGC~5653           & IRASF~14280+3126 & 3562  & 54.9 & 11.06  &H\,{\sc ii}   &1 & isolated\\
NGC~5734           & IRASF~14423$-$2039 & 4074  & 59.3 & 11.06: &NO     &1 &
paired with NGC~5743\\
IC~4518E/W         & IRASF~14544$-$4255 & 4715  & 69.9 & 11.13  &Sy2 (W)  &2 &
close interacting pair \\
Zw~049.057         & IRASF~15107+0724 & 3897  & 59.1 & 11.27: &H\,{\sc ii}
&1,5 & isolated\\
NGC~5936           & IRASF~15276+1309 & 4004  & 60.8 & 11.07  &H\,{\sc ii}   &1 & isolated\\
---                & IRASF~17138$-$1017 & 5197  & 75.8 & 11.42  &H\,{\sc ii}   &2,3 & isolated?\\
IC~4687/IC~4686         & IRASF~18093$-$5744 & 5200/4948  & 74.1 & 11.55:  &H\,{\sc ii} (both)  &3,9 &
close interacting pair\\ 
IC~4734            & IRASF~18341$-$5732 & 4680  & 68.6 & 11.30  &H\,{\sc ii}/L &2 & in group\\
NGC~6701           & IRASF~18425+6036 & 3965  & 56.6 & 11.05  &H\,{\sc ii}:    &1 & isolated\\
NGC~7130           & IRASF~21453$-$3511 & 4842  & 66.0 & 11.35  &L/Sy  &1,2 & peculiar\\
IC~5179            & IRASF~22132$-$3705 & 3422  & 46.7 & 11.16  &H\,{\sc ii}   &1 & isolated\\
NGC~7469           & IRASF~23007+0836 & 4892  & 65.2 & 11.59  &Sy1   &1 &
paired with IC~5283\\
NGC~7591           & IRASF~23157+0618 & 4956  & 65.5 & 11.05  &L     &1 & isolated\\
NGC~7771           & IRASF~23488+1949 & 4277  & 57.1 & 11.34   &H\,{\sc ii}   &1&
paired with NGC~7770\\
\enddata
\tablecomments{Column~(1): Galaxy name. Column~(2): {\it IRAS} denomination
  from Sanders et al. (2003). Column~(3): 
Heliocentric velocity from NED. Column~(4): Distance taken from Sanders et al. (2003).
Column~(5): $8-1000\,\mu$m IR
luminosity taken from Sanders et al. (2003), where the suffix ``:'' means
large uncertainty (see Sanders et al. 2003 for details). NGC~633 was
identified as the {\it IRAS} source in Sanders et al. (1995).
Interacting pairs for which Surace et al. (2004) obtained {\it IRAS} fluxes for
the individual components: NGC~633 $\log L_{\rm IR}=10.56$, and for NGC~2388,
NGC~7469, and NGC~7771 with values of $\log
L_{\rm IR}$ similar to those given in Sanders et al. (2003). 
Column~(6): Nuclear activity class 
from spectroscopic data (``Sy''=Seyfert, ``L''=LINER, ``H\,{\sc ii}''=H\,{\sc ii}
region-like, ``NO''=no classification was possible, ``--''=no data
available). Column~(7): References for the spectroscopic data: 
1. Veilleux et al. (1995);
2. Corbett et al. (2003);
3. Kewley et al. (2001); 
4. V\'eron-Cetty \& V\'eron  (2001);
5. Baan, Salzer, \& LeWinter (1998); 
6. L\'{\i}pari et al. (2000);
7. Garc\'{\i}a-Mar\'{\i}n et al. (2006). The Sy2 classification is for B1 in NGC~3690;
8. van den Broek  et al. (1991);
9. Sekiguchi \& Wolstencroft (1992).  
Column~(8): Morphological description.}

\end{deluxetable*}


In this paper we present {\it Hubble Space Telescope} ({\it HST}) 
NICMOS continuum (1.1, 1.6, and
$1.876\,\mu$m) and Pa$\alpha$ emission line observations of a volume limited
sample (30 systems, 34 individual galaxies) 
of local universe ($z<0.017$) LIRGs.  
The paper is organized as follows. \S2 gives details on the sample
selection and properties. In \S3 we describe the {\it  HST}/NICMOS observations, data
reduction, and continuum and H\,{\sc ii} region photometry. In \S4 we discuss
the near-IR continuum properties of the central regions of local LIRGs. In
\S5, and \S6,we
describe the Pa$\alpha$ morphology, and the properties of individual H\,{\sc
  ii} regions, respectively. In \S7 we estimate the extinction to the Pa$\alpha$
emitting regions. In \S8  we estimate the extended H$\alpha$ emission, and in 
\S9 we discuss the star formation rates of local LIRGs. Our conclusions are given in \S10.
Throughout this paper we use $H_0=75\,$km s$^{-1}$ Mpc$^{-1}$.

\section{The sample}
We have selected a volume limited sample of nearby LIRGs from the {\it IRAS}
Revised Bright 
Galaxy Sample (RBGS: Sanders et al. 2003). The velocity 
range ($v_{\rm hel}=2750-5200\,{\rm km\,s}^{-1}$) of the galaxies was chosen 
so that the Pa$\alpha$ emission line ($\lambda_{\rm rest} =
1.876\,\mu$m) falls into the {\it HST}/NICMOS F190N narrow-band filter. We required
the logarithm of the total IR luminosity to be $\ge$ 11.05 and that the
galaxies be at Galactic latitude $|b| > 10\deg$. The sample 
is composed of 30 systems (34 individual galaxies) and contains 77\% of the
complete sample of systems within this velocity range in 
the RBGS. 

The sample is presented in Table~1. The IR luminosities (and distances) 
of the systems are
taken from the  RBGS 
(Sanders et al. 2003), and are in the range 
 $\log L_{\rm IR}  = 11.05-11.88\,[$L$_\odot]$. 
Five (NGC~1614, NGC~3256, NGC~3690/IC~694, 
Zw~049.057, and NGC~7469) of these systems were part of
the NICMOS guaranteed time observations 
(GTO, see Scoville et al. 2000; Alonso-Herrero et al. 
2000, 2001, 2002),  and one (NGC~5653) was part of the 
NICMOS Snapshot Survey of Nearby Galaxies (B\"oker et al. 1999).
The remaining twenty-four systems are new general observer (GO) observations
obtained during {\it HST} Cycle 13.

In Table~1 we also list, if available, the nuclear activity 
class --- H\,{\sc ii}, Seyfert and LINER --- obtained
from spectroscopic observations in the literature (see Table~1 notes for
references). The 
galaxies show a variety of morphologies and environments, 
including isolated systems, galaxies in groups, 
interacting galaxies, and advanced mergers. In  the case of close interacting
galaxies the listed IR luminosities correspond to both
galaxies in the system. 
Surace, Sanders, \& Mazzarella (2004) presented high resolution
{\it IRAS} images for close interacting systems within the RBGS. Their work
contains four systems in our sample for which the {\it IRAS} fluxes of the individual
components of the pair are available, as listed in the notes to Table~1. There are four
further systems for which Surace et al. (2004) could not obtain {\it IRAS}
fluxes for the two individual components: MCG~$-$02-33-098E/W, IC~4518E/W,
IC~4686/IC~4687, and Arp~299. The components of Arp~299 have been amply studied 
in the mid-IR and their contributions to the total IR luminosity of the system
are well determined (see e.g., Soifer et al. 2001).

For various reasons, nine galaxies that meet our selection criteria have not
been imaged in Pa$\alpha$: UGC~2982, CGCG~468-002, ESO~264-G057, MCG~$-$03-34-064, 
IC~4280, UGC~8739, ESO~221-G010, NGC~5990, and NGC~7679. We have examined 
these galaxies carefully to see if their properties depart from
those in the observed sample. They do not. For example, the average logarithm of the luminosity
of the observed galaxies is $\log L_ {\rm IR} = 11.25\,[{\rm L}_\odot]$, 
while for the unobserved ones it is $\log L_ {\rm IR} = 11.11\,[{\rm L}_\odot]$. The average 
redshift of the former group is 4317 km s$^{-1}$, and for the latter is 4722
km s$^{-1}$. The IR spectral shapes are generally similar, except that the observed
sample contains both of the galaxies that {\it IRAS} did not detect at $12\,\mu$m: IC~860
and Zw~049.057.

Within the errors, these galaxies all have approximately solar metallicity. Although
this conclusion is expected, we were able to confirm it for all but three members
of the sample (MCG~+12-02-001, UGC~1845, and IC~860). We did so from measurements
of [N\,{\sc ii}]/H$\alpha$ and [O\,{\sc iii}]/H$\beta$ from the literature, converted to
$12 + \log{\rm (O/H)}$ according to the correlations found by 
Melbourne \& Salzer (2002). For the first two exceptions, we could not find adequate data, 
while for IC~860 the emission line equivalent width is very small and the
data in the literature are not consistent. For all the others, the computed
metallicity is within a factor of two of solar. Given the approximate 
approach to the calculations, this factor is easily within the errors.   

For our selected range of distance, the {\it IRAS} sample of LIRGs should be essentially
complete (excluding regions of high IR cirrus and those not surveyed). Our
observed sample is 77\% complete in terms of the IRAS sample, and the missing
galaxies are very similar to those observed. We conclude that our results will
be representative of local LIRGs in general.

\section{{\it HST}/NICMOS Observations}

In this section we only describe the new {\it HST}/NICMOS observations obtained in Cycle 13. 
We refer the reader to Scoville et al. (2000)
and Alonso-Herrero et al. (2000, 2001, 2002) for details on the GTO observations,
data reduction, and flux calibration of the data obtained prior to Cycle 13
(see \S2).

\subsection{Data reduction}

The observations of the LIRGS were taken with the NIC2 camera 
(plate scale of 0.076\arcsec pixel$^{-1}$) using the F110W and
F160W broad-band filters, and the F187N and F190N 
narrow-band filters. At the distances of the LIRG sample, 
the F190N filter contains the Pa$\alpha$ emission line  and the adjacent
continuum at $1.90\,\mu$m.  For  continuum subtraction, we used the images
taken through the  F187N filter. The field of view (FOV) of the images is 
approximately $19.5\arcsec \times 19.5\arcsec$ which for our sample covers
the central $3.3-7.2\,$kpc of the galaxies. The observations were designed such that each galaxy
could be observed within one {\it HST} orbit. The typical integration times
were 250\,s for each of the broad-band filters and $900-950$\,s for each of the narrow-band filters.
For each filter the total integration time was split into three individual
exposures with relative offsets of 5 pixels.

The images were reduced using the {\it HST}/NICMOS pipeline
routines, which involve subtraction of the first readout, 
dark current subtraction on a readout-by-readout basis, correction
for linearity and cosmic ray rejection, and flat fielding. 
For a given filter, the individual images were aligned using common features
present in all three images and combined to construct a mosaic. 
The final F110W images,  continuum-subtracted
Pa$\alpha$ images, and ${\rm F110W}-{\rm F160W}$ color maps are presented in
Fig.~1 for the new observations of galaxies in our sample.

The flux calibration of the images was performed using
conversion factors given in the {\it HST}/NICMOS handbook.
We have also examined the near-IR continuum images and the 
$m_{\rm F110W}-m_{\rm F160W}$ color maps (Fig.~1) to look for 
the presence of nuclear point sources. When a nuclear point source was
detected we obtained photometry 
with a 1\arcsec-diameter aperture (subtracting the underlying galaxy emission measured
in an annulus around the point source) 
and performed an aperture correction 
(from the NICMOS handbook) 
to include all the flux
from the point source. The typical photometry uncertainties  associated with removing the
underlying galaxy emission were $0.04-0.06\,$mag. If a nuclear point source was not detected, we measured the
continuum emission using a similarly sized aperture but without removing the
emission from the underlying galaxy. The results for the nuclei are
presented in Table~2, and discussed in \S4.2.


\begin{figure*}
\vspace{0.5cm}

\includegraphics[angle=-90,width=15cm]{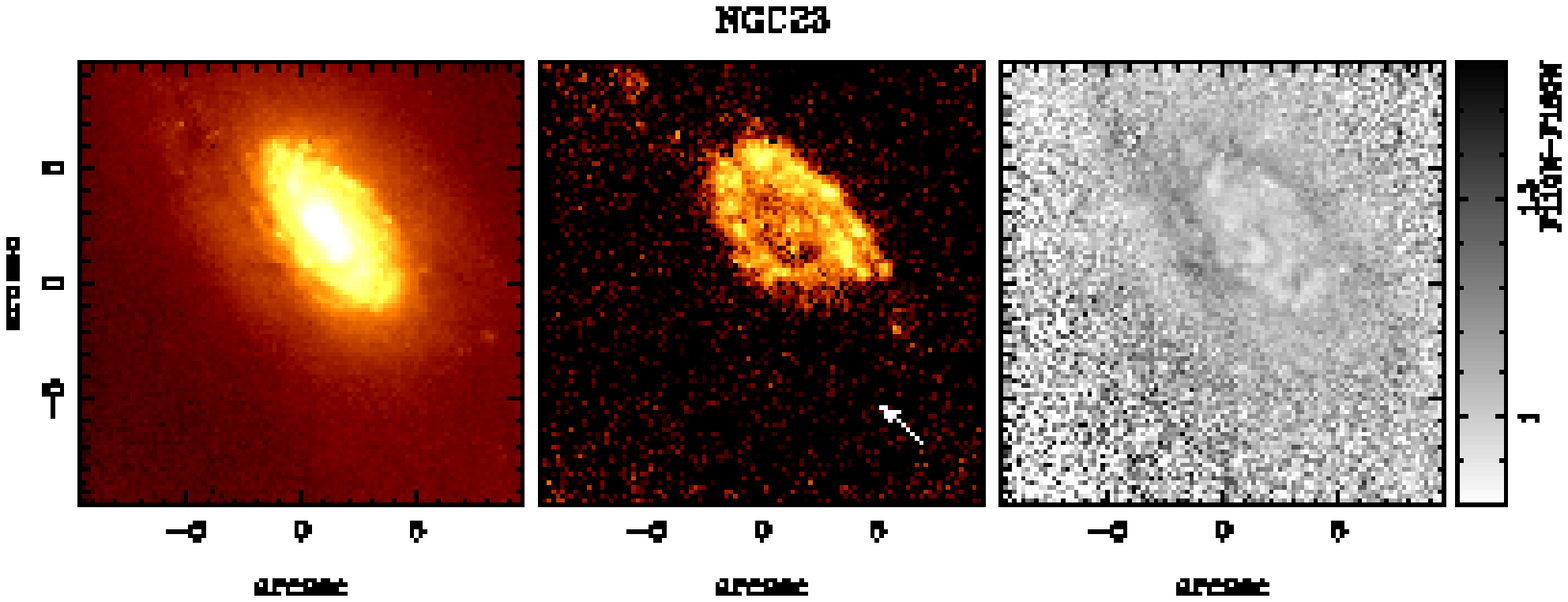}

\vspace{0.5cm}

\includegraphics[angle=-90,width=15cm]{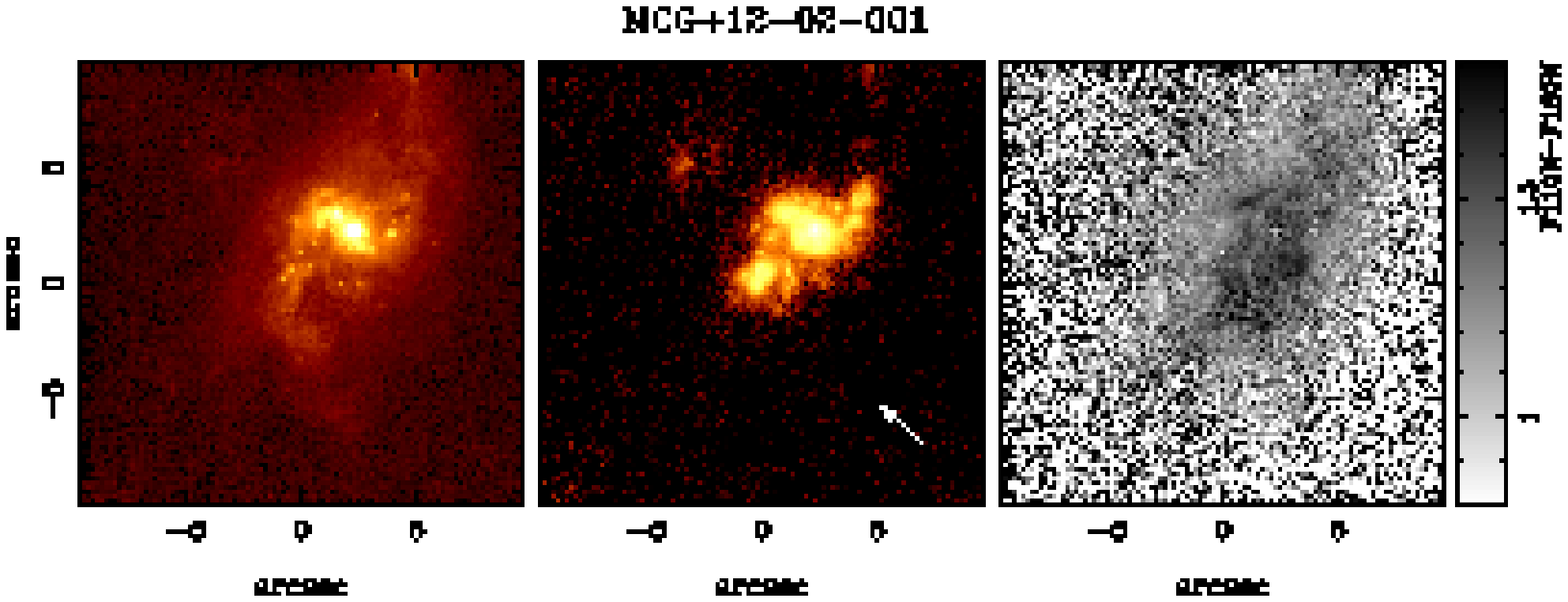}

\vspace{0.5cm}

\includegraphics[angle=-90,width=15cm]{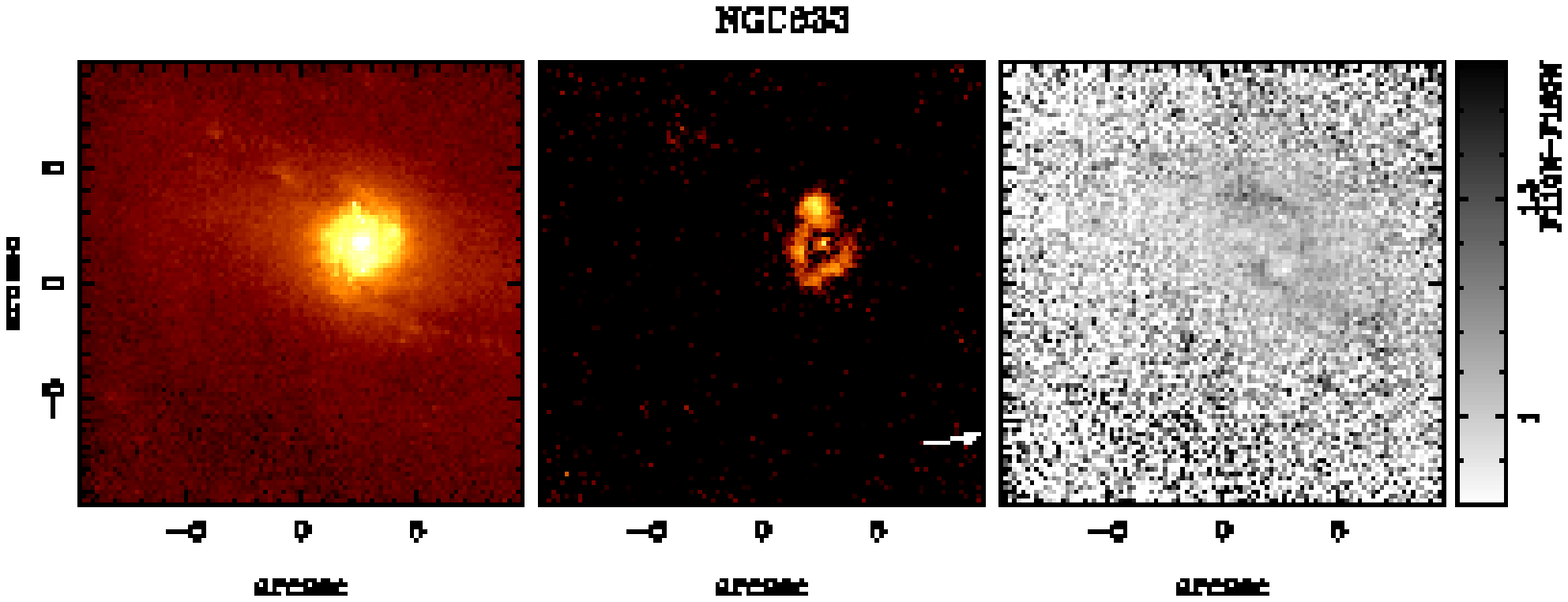}
\vspace{0.5cm}

\includegraphics[angle=-90,width=15cm]{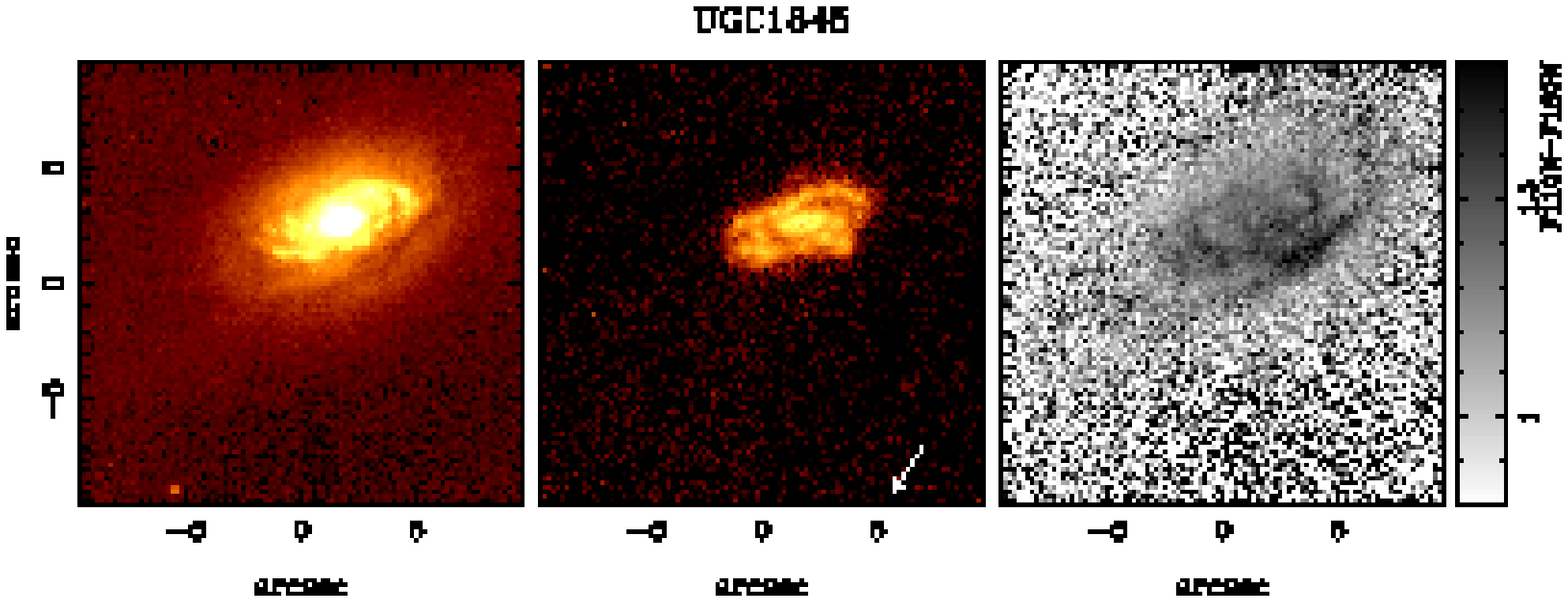}

\caption{{\it HST}/NICMOS $1.1\,\mu$m continuum emission images (F110W filter, left panels), continuum-subtracted 
Pa$\alpha$ line emission images (F190N-F187N filters, 
middle panels), and $m_{\rm F110W} - m_{\rm F160W}$
color maps (right panels, the scale on the right hand side of the images is
the $m_{\rm F110W} - m_{\rm F160W}$ color in
magnitudes) for the galaxies in our sample observed during Cycle
13.  The arrow indicates north, and east is in the counterclockwise direction.}
\end{figure*}

\begin{figure*}
\setcounter{figure}{0}

\vspace{0.5cm}

\includegraphics[angle=-90,width=15cm]{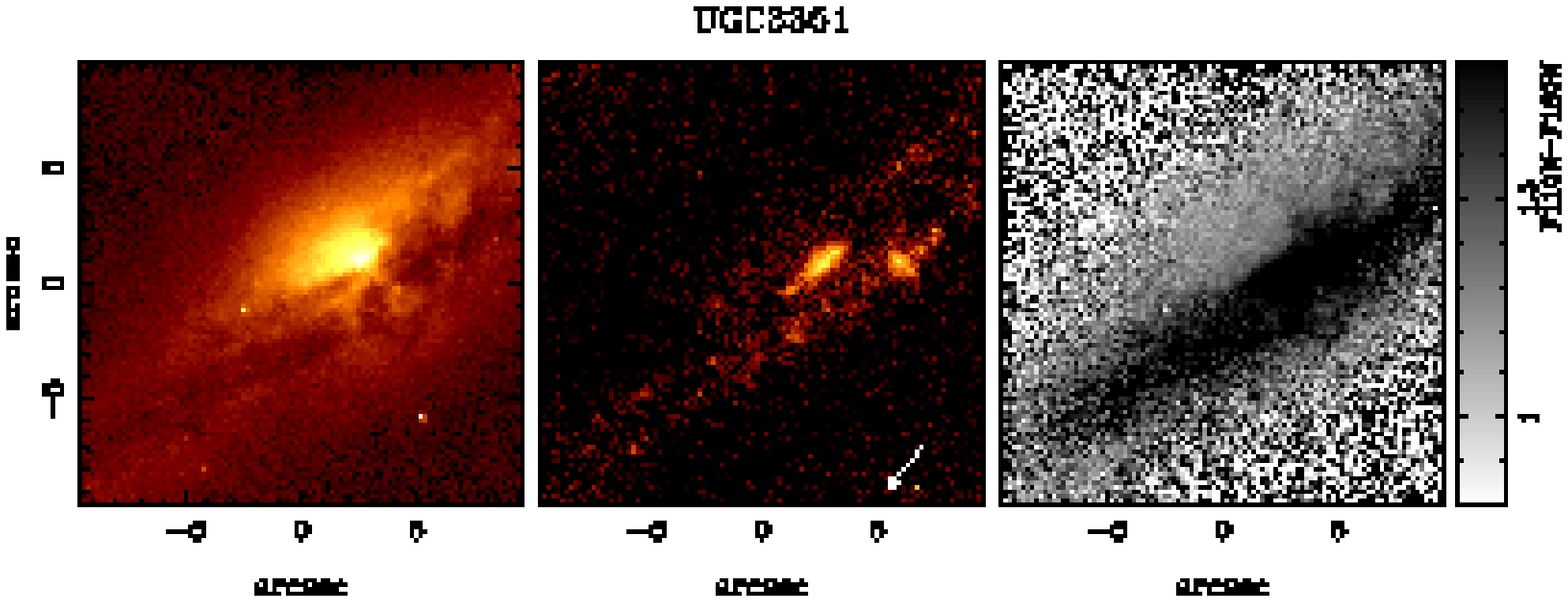}

\vspace{0.5cm}

\includegraphics[angle=-90,width=15cm]{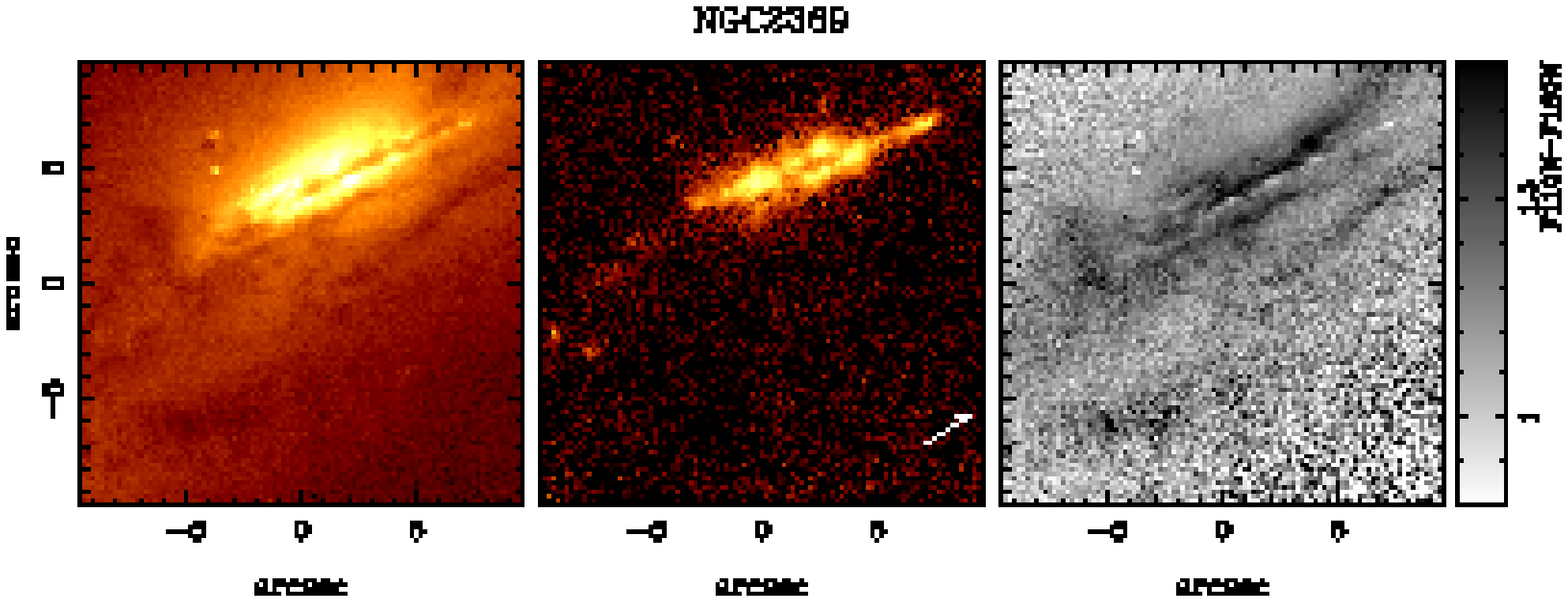}

\vspace{0.5cm}

\includegraphics[angle=-90,width=15cm]{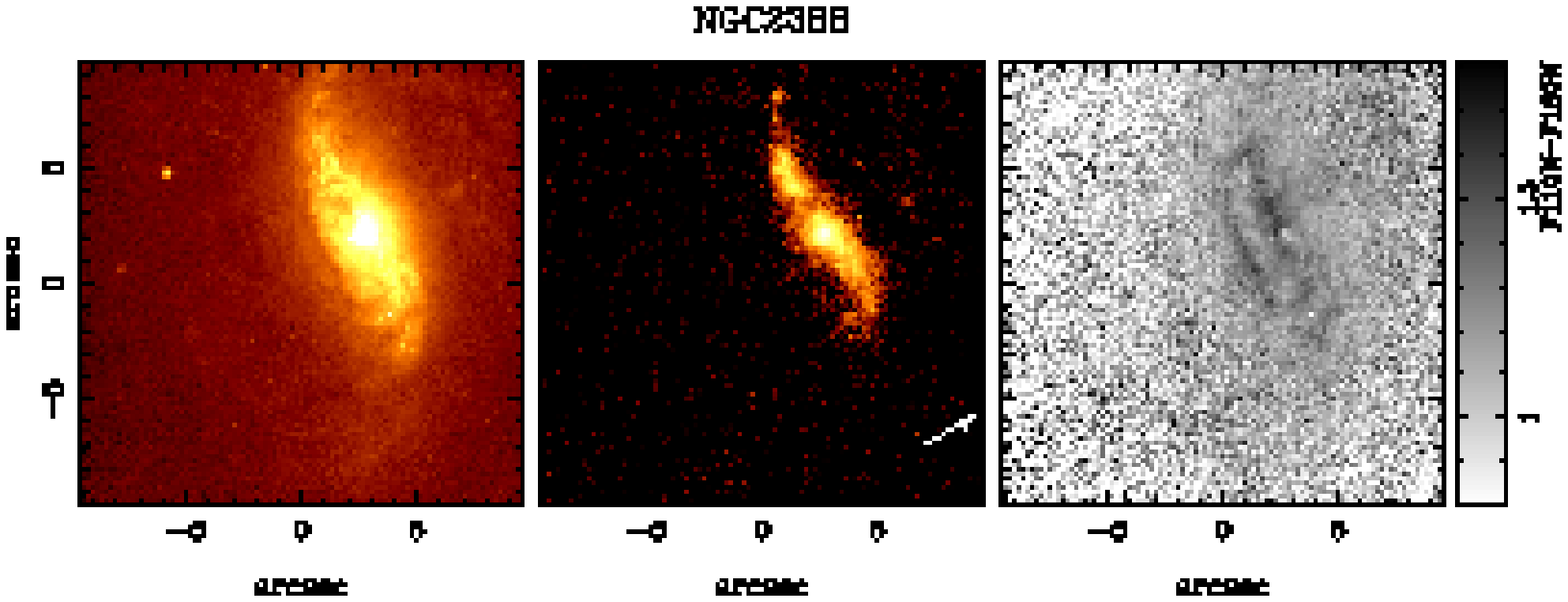}

\vspace{0.5cm}

\includegraphics[angle=-90,width=15cm]{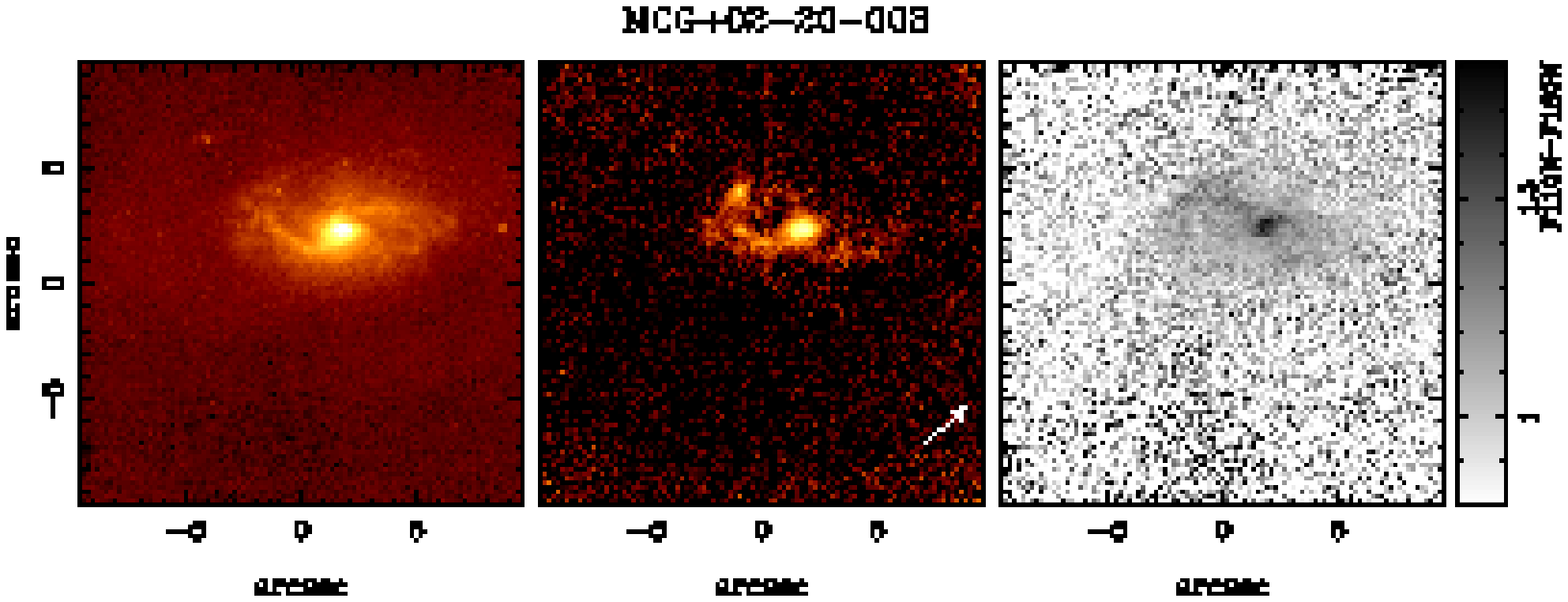}

\caption{Continued.}
\end{figure*}

\begin{figure*}
\setcounter{figure}{0}

\vspace{0.5cm}

\includegraphics[angle=-90,width=15cm]{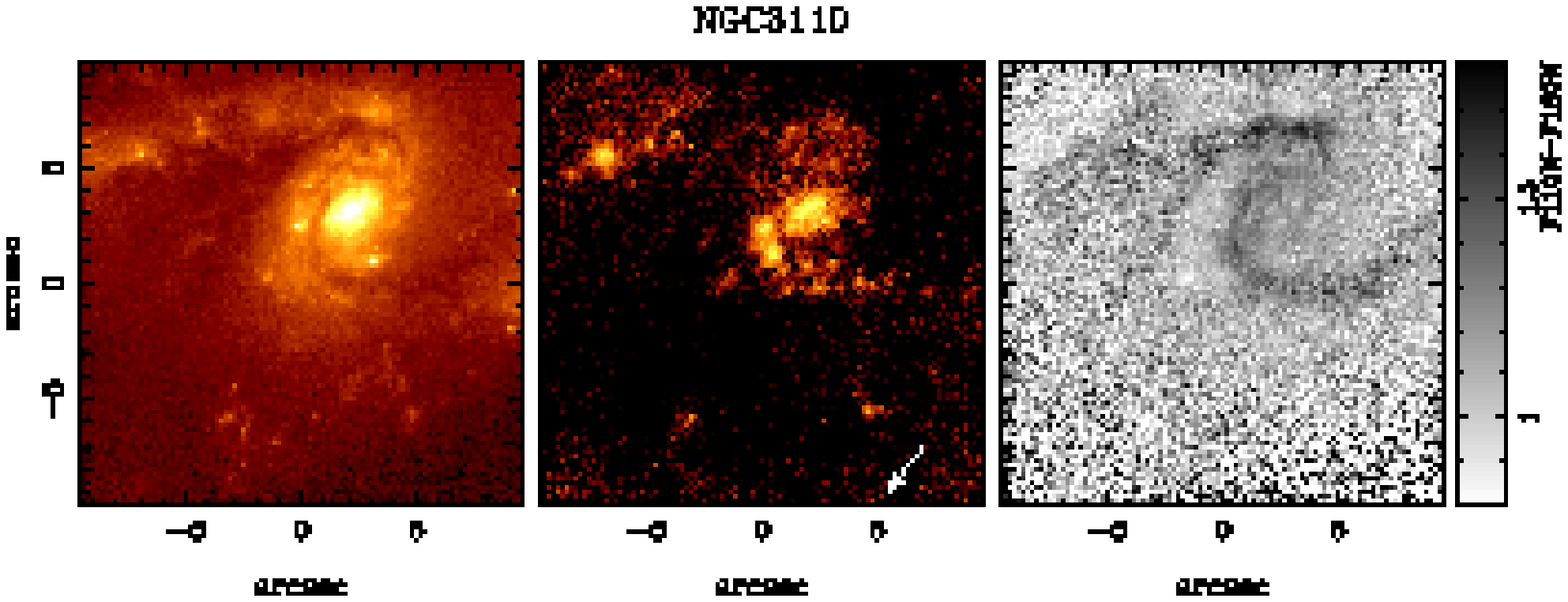}

\vspace{0.5cm}

\includegraphics[angle=-90,width=15cm]{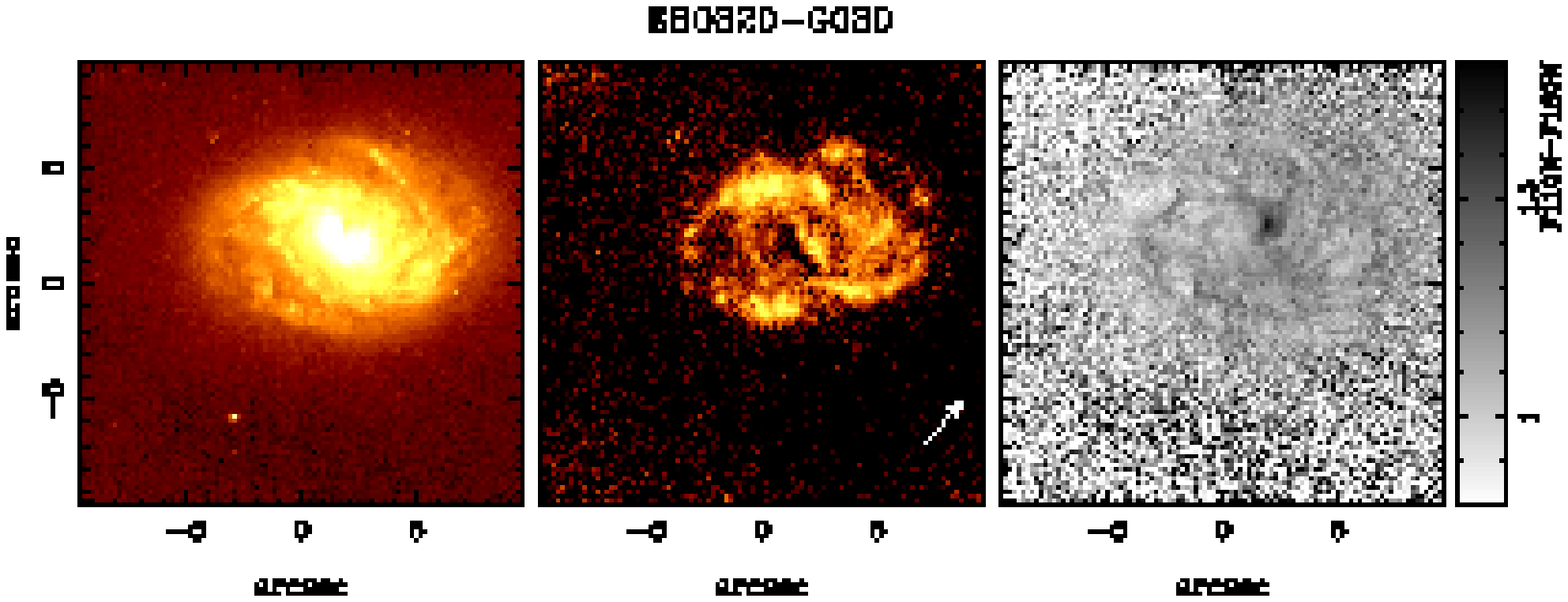}

\vspace{0.5cm}

\includegraphics[angle=-90,width=15cm]{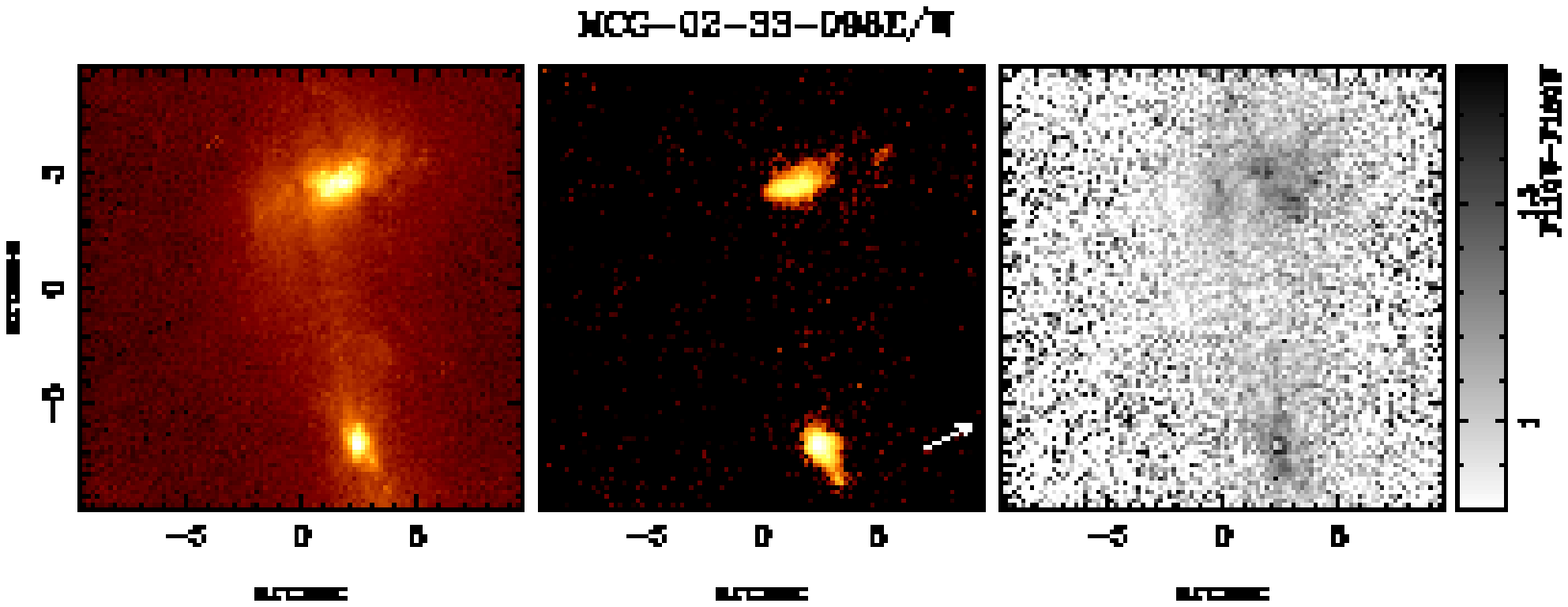}

\vspace{0.5cm}

\includegraphics[angle=-90,width=15cm]{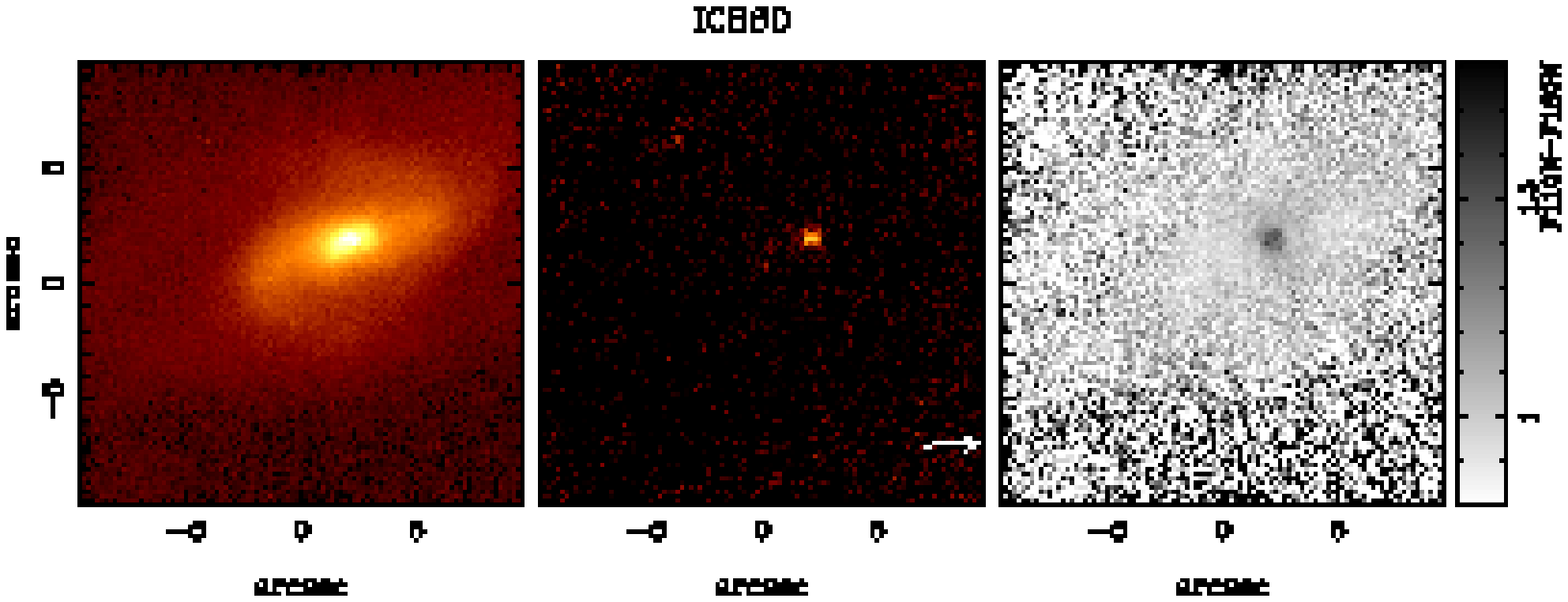}

\caption{Continued.}
\end{figure*}


\begin{figure*}
\setcounter{figure}{0}

\vspace{0.5cm}

\includegraphics[angle=-90,width=15cm]{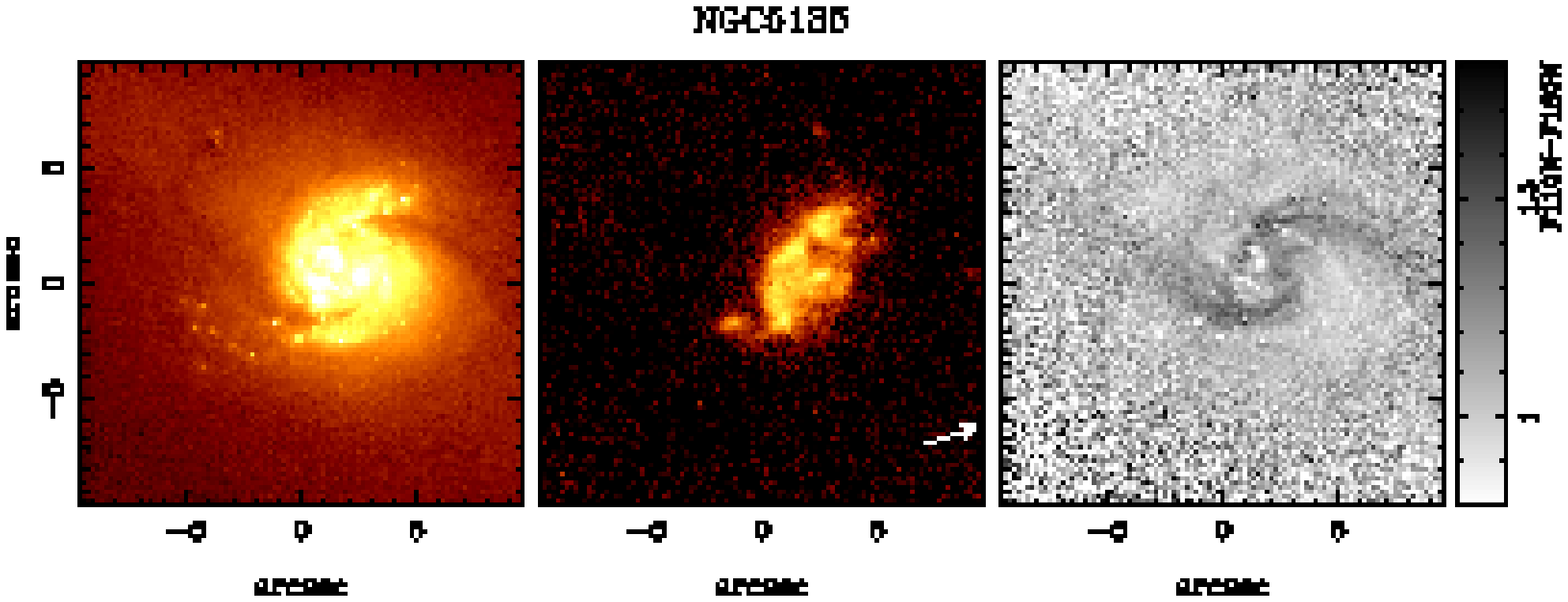}

\vspace{0.5cm}

\includegraphics[angle=-90,width=15cm]{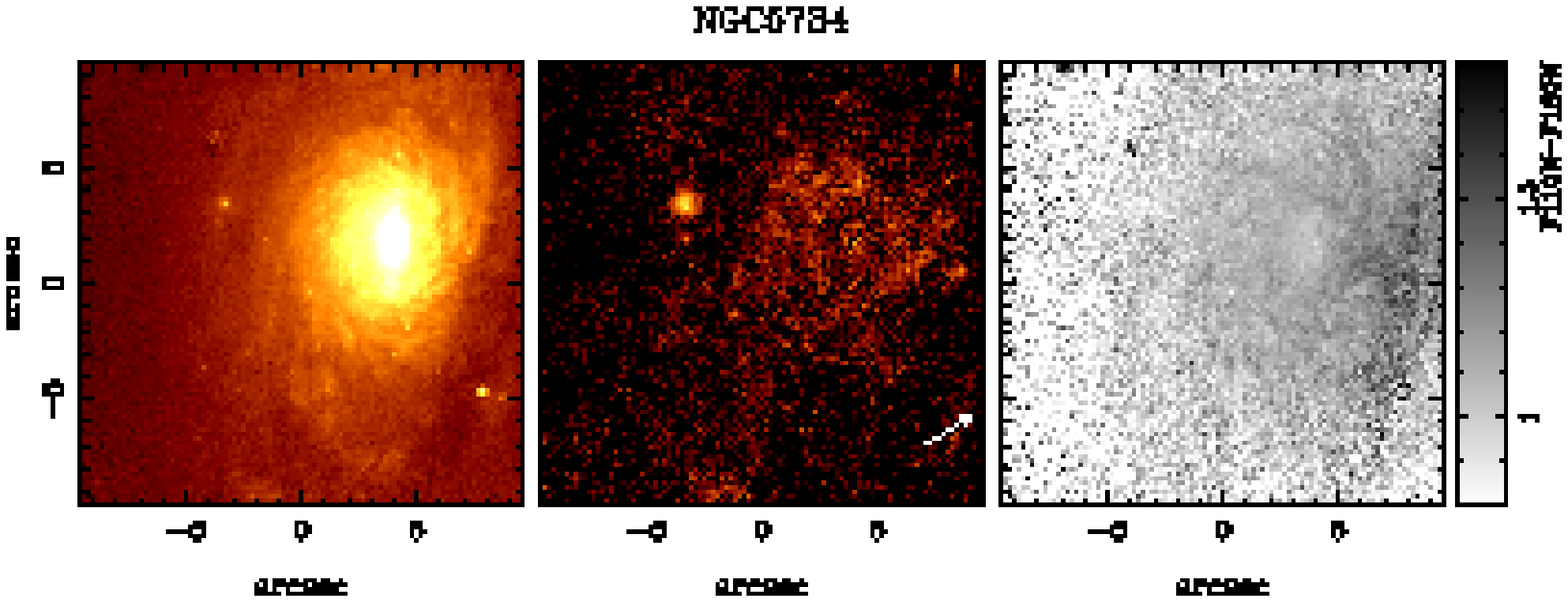}

\vspace{0.5cm}

\includegraphics[angle=-90,width=15cm]{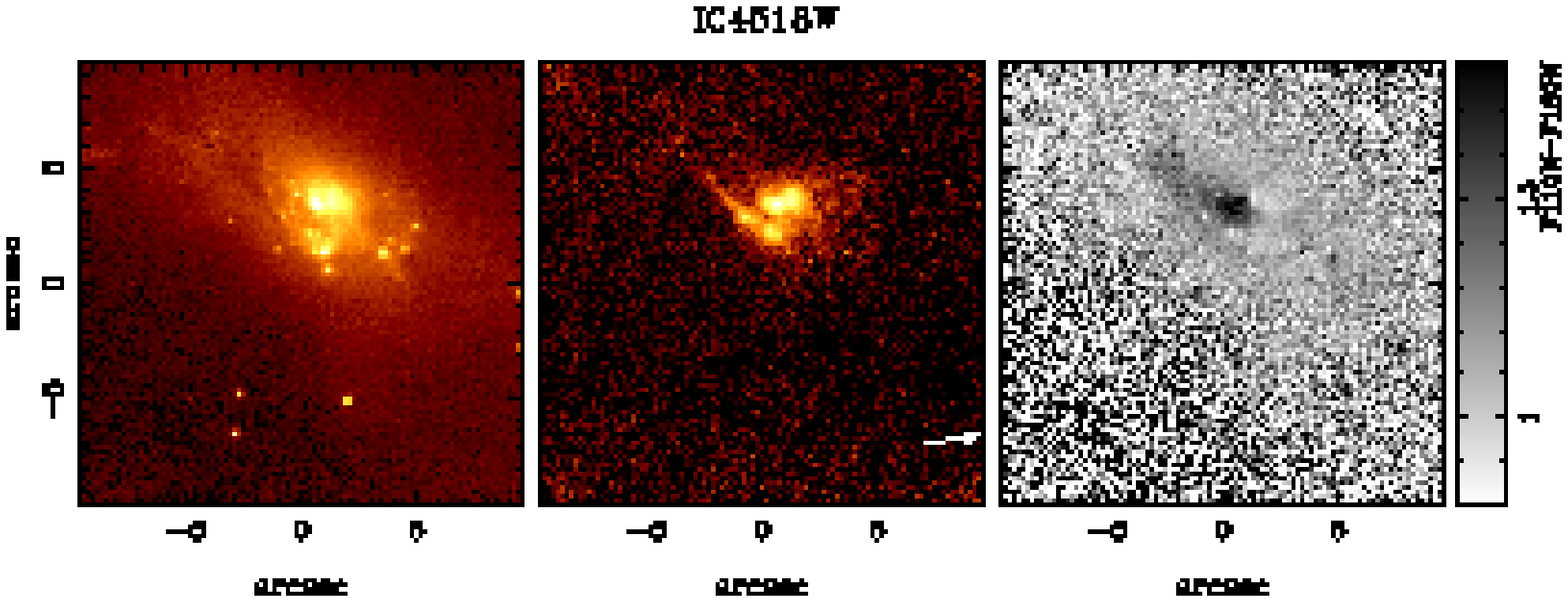}

\vspace{0.5cm}

\includegraphics[angle=-90,width=15cm]{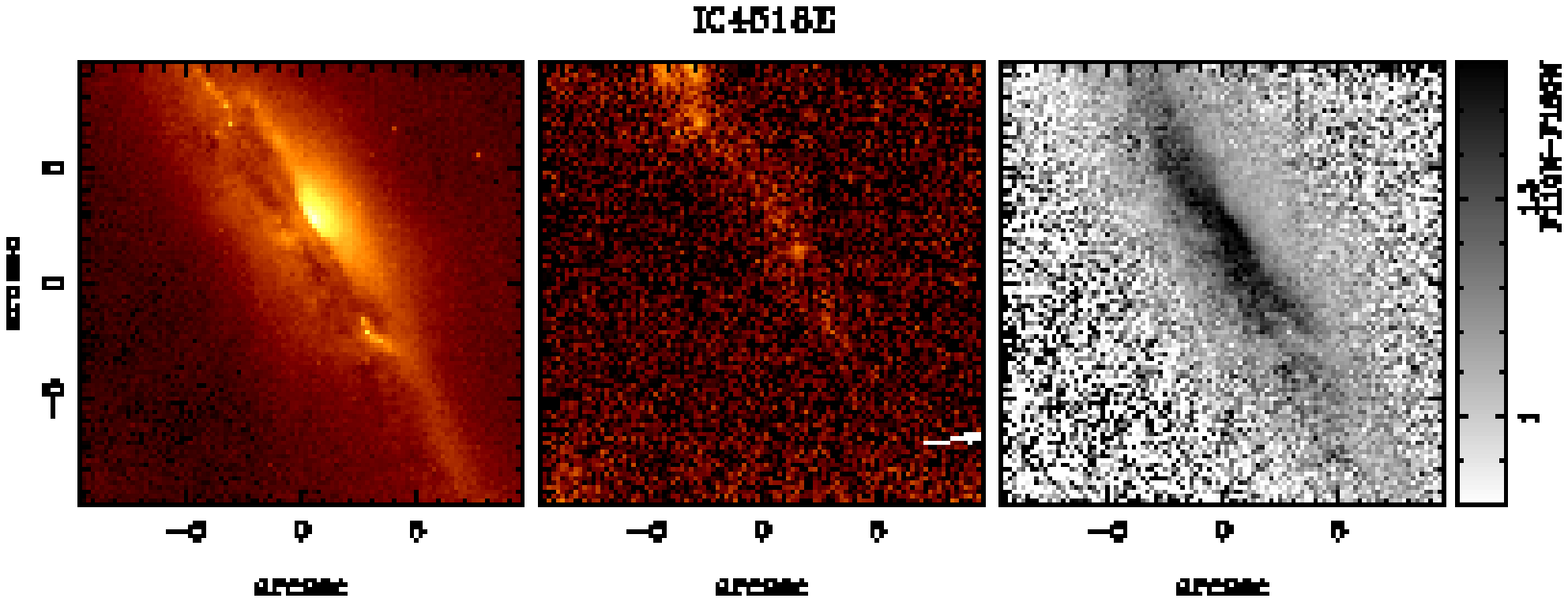}

\vspace{0.5cm}

\caption{Continued.}
\end{figure*}

\begin{figure*}
\setcounter{figure}{0}

\includegraphics[angle=-90,width=15cm]{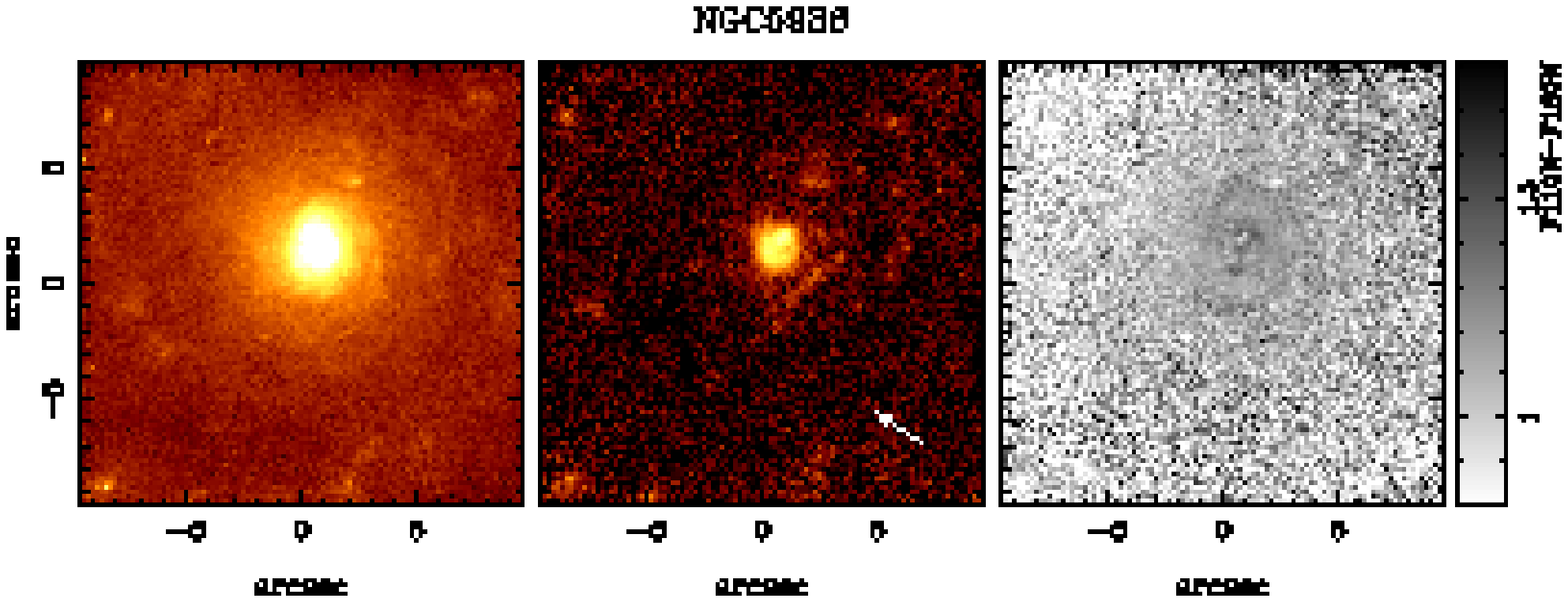}

\vspace{0.5cm}

\includegraphics[angle=-90,width=15cm]{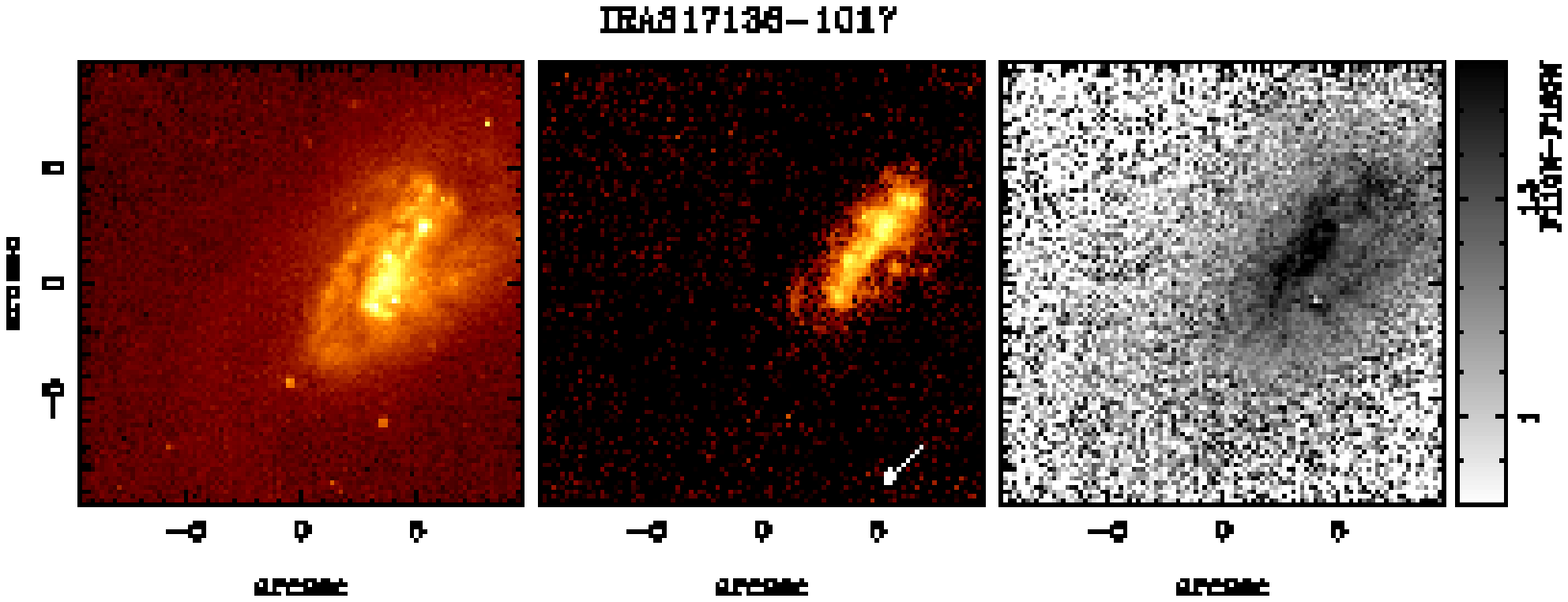}

\space{0.5cm}

\includegraphics[angle=-90,width=15cm]{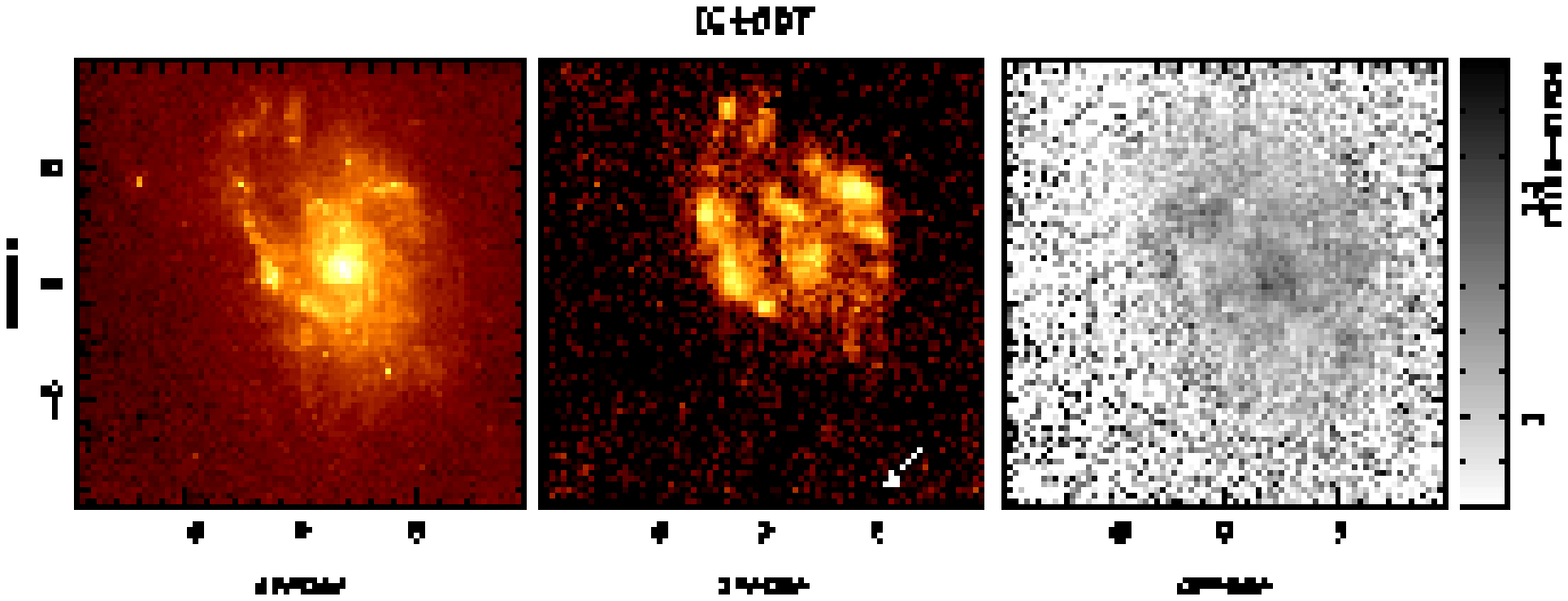}

\vspace{0.5cm}

\includegraphics[angle=-90,width=15cm]{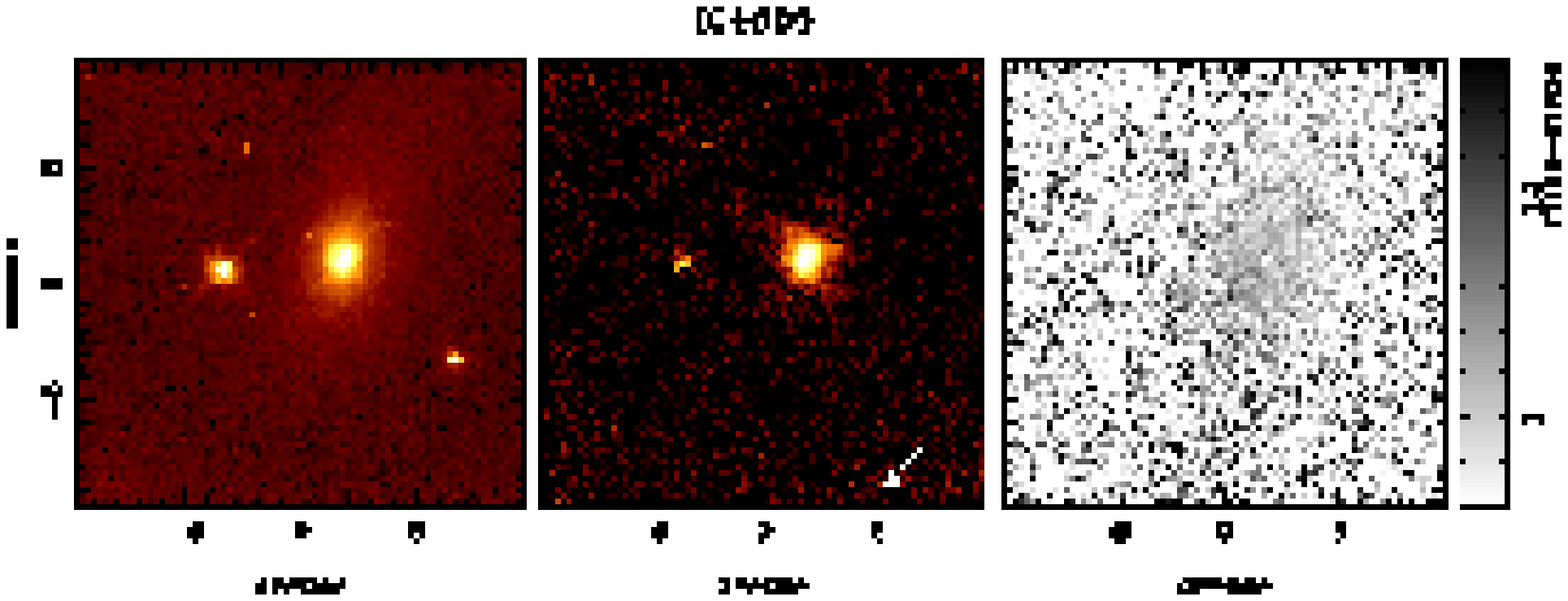}


\caption{Continued.}
\end{figure*}
 
\begin{figure*}
\setcounter{figure}{0}

\vspace{0.5cm}

\includegraphics[angle=-90,width=15cm]{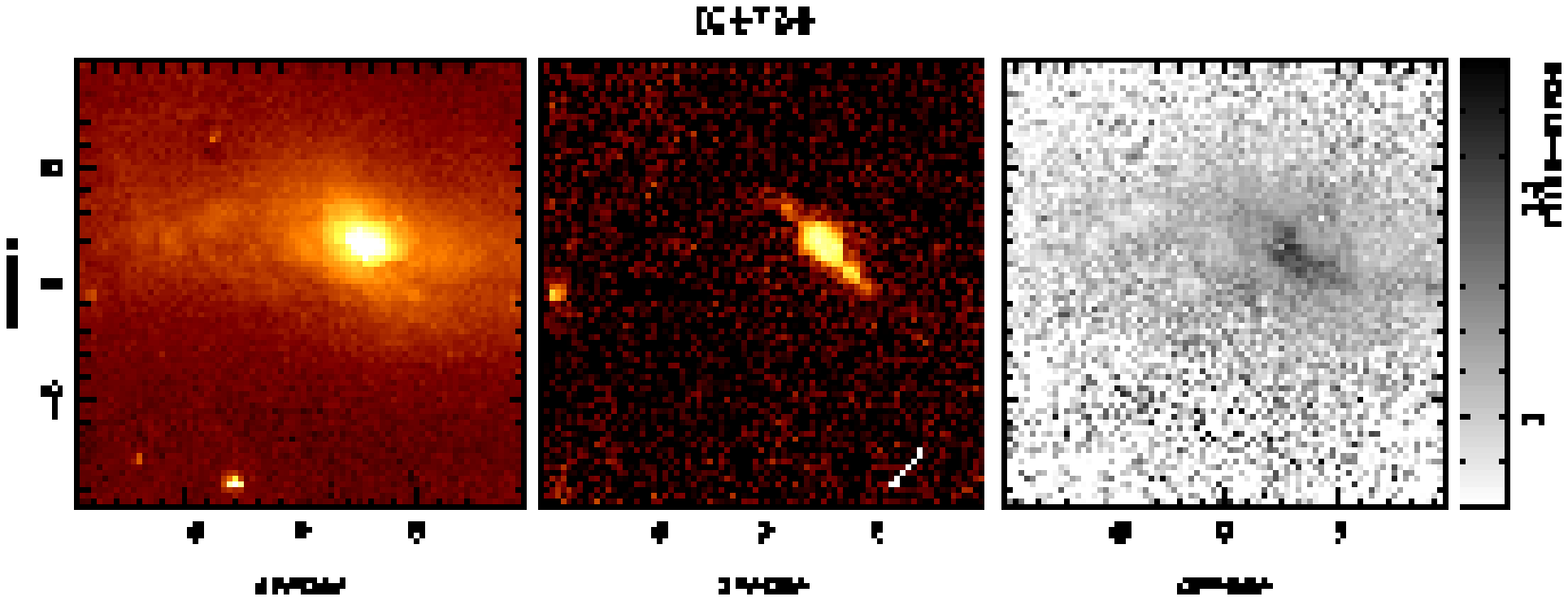}

\vspace{0.5cm}

\includegraphics[angle=-90,width=15cm]{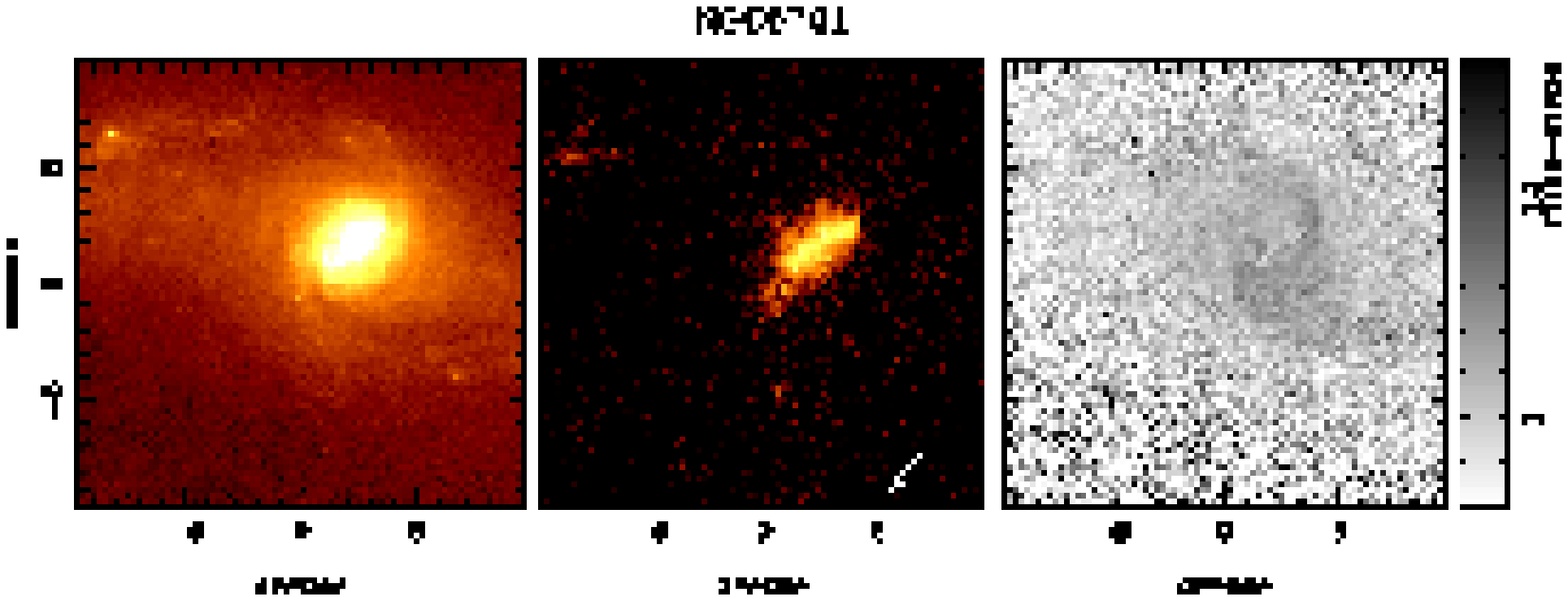}

\vspace{0.5cm}

\includegraphics[angle=-90,width=15cm]{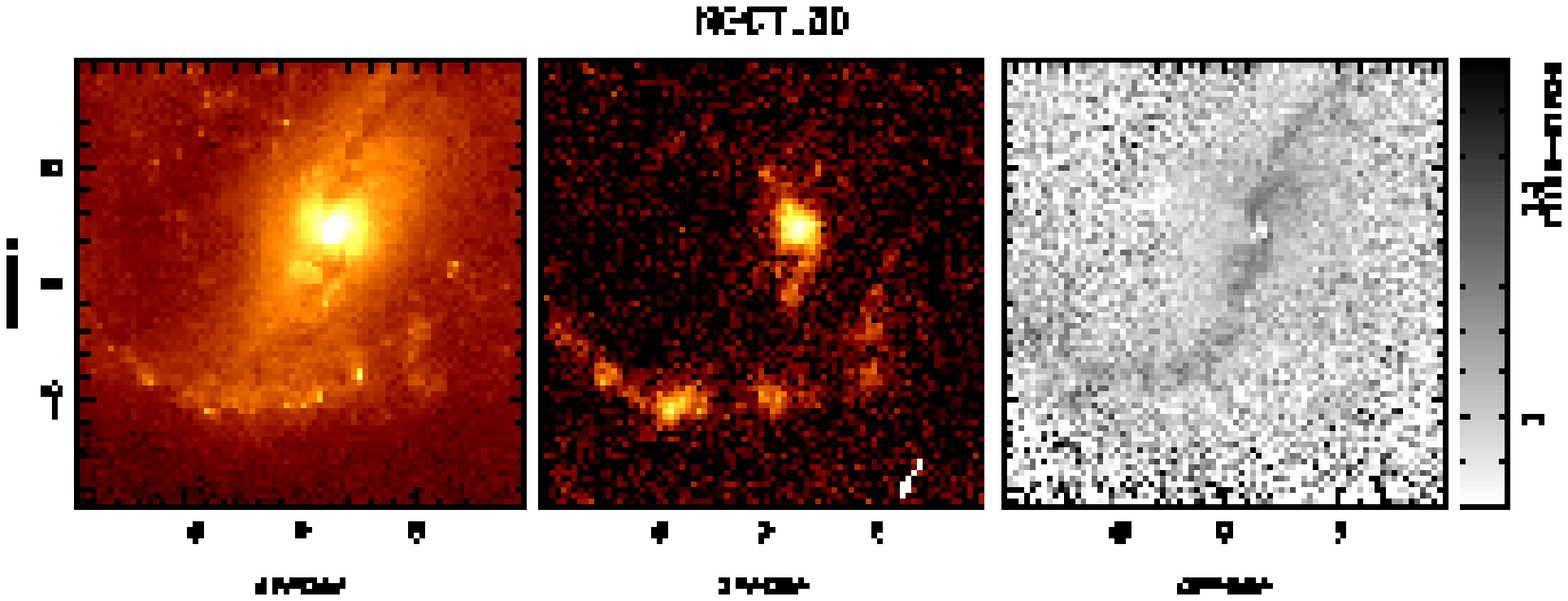}

\vspace{0.5cm}

\includegraphics[angle=-90,width=15cm]{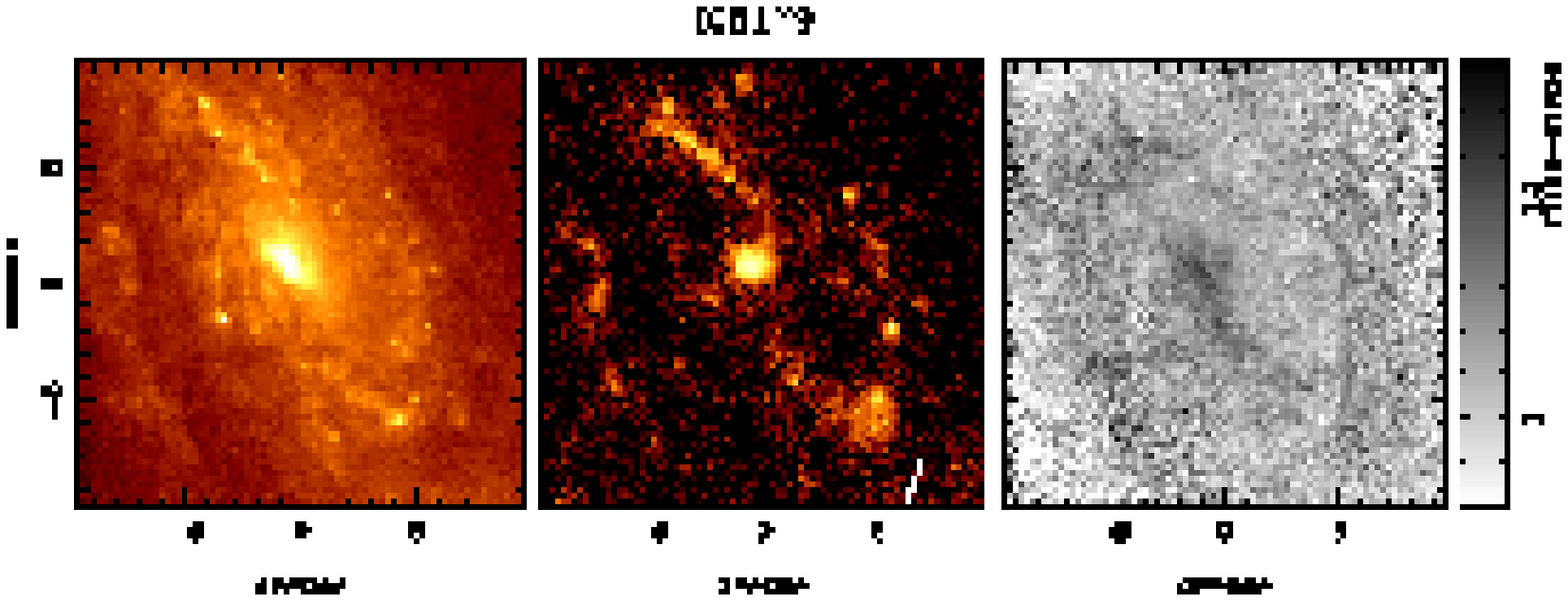}

\caption{Continued.}
\end{figure*}


\begin{figure*}
\setcounter{figure}{0}
\vspace{0.5cm}

\includegraphics[angle=-90,width=15cm]{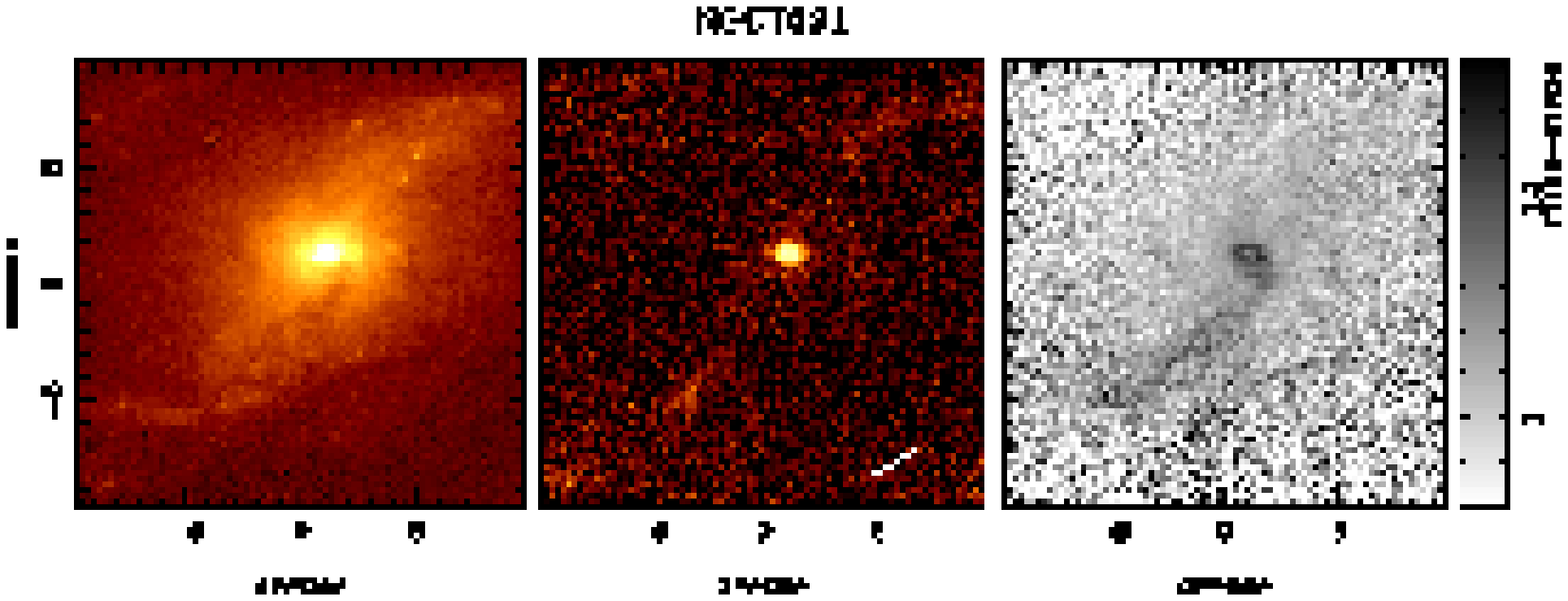}

\vspace{0.5cm}

\includegraphics[angle=-90,width=15cm]{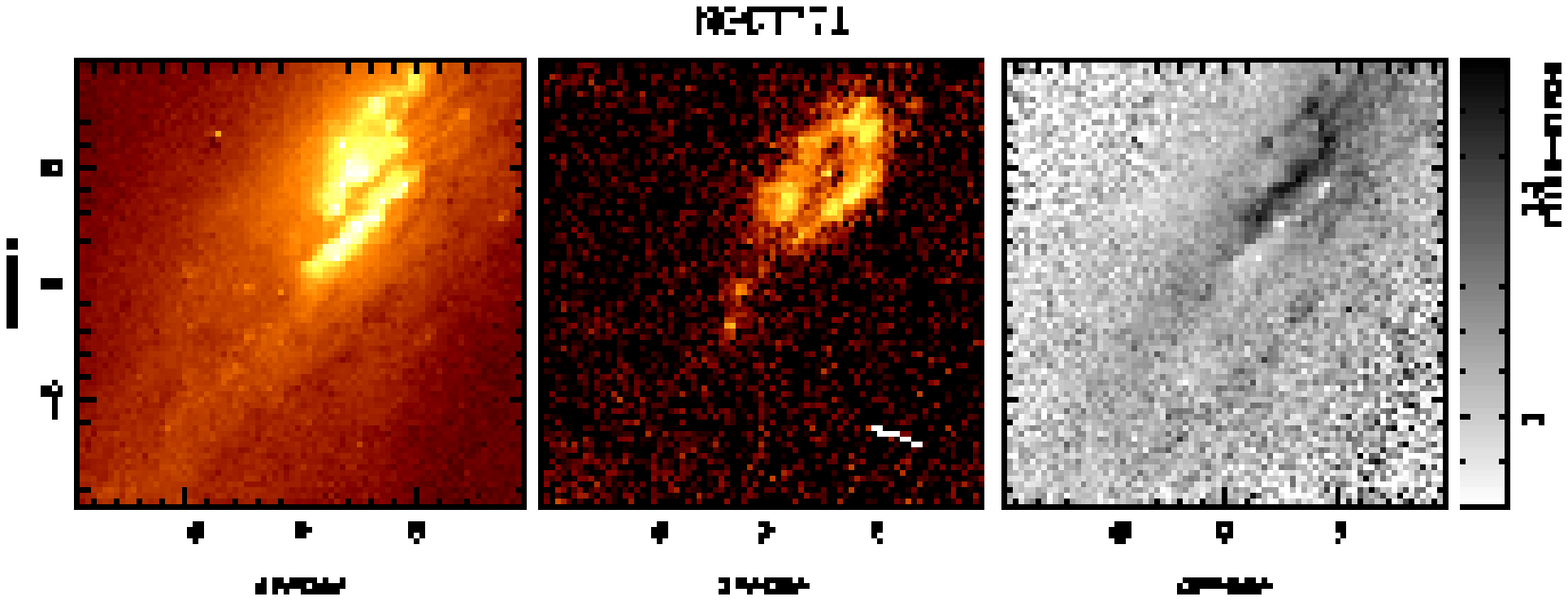}

\caption{Continued.}
\end{figure*}

\subsection{Photometry of H\,{\sc ii} regions}
The continuum-subtracted Pa$\alpha$ line emission images were produced by
subtracting the F187N images from the F190N images, both calibrated in Jy.
Using the F190N filter we estimate that for galaxies at 
$5000 \lesssim v \lesssim 5200\,{\rm km \, s}^{-1}$ up to 20\% (for
the typical widths of hydrogen recombination 
lines of LIRGs not containing a type 1 AGN, see Goldader et al. 1997a) of
the Pa$\alpha$ flux could be lost due to
the observed wavelength of the emission line (see discussion in Alonso-Herrero
et al. 2002).

We have used SE{\sc xtractor} (Bertin \& Arnouts 1996) to analyze the 
properties of
the individual H\,{\sc ii} regions detected in the Pa$\alpha$ images of our
sample of galaxies. We have set the lower limit for the size of an H\,{\sc ii}
region to 9 contiguous pixels. For the distances of the galaxies 
in our sample, this corresponds to minimum linear sizes (equivalent diameters) of between 
44 and 96\,pc. The detection threshold criterion 
for a pixel to be included as part of an H\,{\sc ii} region is to be above the
background plus twice the rms noise of the
background. SE{\sc xtractor} constructs a background image by a bi-cubic
interpolation over areas of the background with a size 
specified by the user. These areas ought 
to be larger than the typical size of an H\,{\sc ii} region, but not so large
as to smooth over local variations of the background. The main
source of uncertainty in measuring the Pa$\alpha$ fluxes of the H\,{\sc ii}
regions is the background removal. After modelling the background and 
 H\,{\sc ii} regions, we output their positions, and
isophotal fluxes, errors, and areas after subtracting the global 
background from the observed Pa$\alpha$ flux. 

The properties of the H\,{\sc ii} regions detected in the LIRGs are summarized
in Table~3. 
We list the H$\alpha$ luminosity (the Pa$\alpha$/H$\alpha$ ratio varies by about 10\% 
for various conditions and under Case B; we take a ratio of 8.6 
from Hummer \& Storey
1987)  for the median  H\,{\sc ii} region, the brightest H\,{\sc ii} region, 
and the average H$\alpha$ luminosity of the three brightest H\,{\sc ii} regions (first-ranked H\,{\sc ii}
regions), as well as the ratio of the sum of the luminosities of the 
three brightest H\,{\sc ii} regions to
the observed H$\alpha$ luminosity in H\,{\sc ii} regions, and the  observed
H$\alpha$ luminosity in H\,{\sc ii} regions.  These properties are discussed in \S6.

\section{Near-IR continuum properties of the central regions of LIRGs}

\subsection{Morphology}

Morphological studies of complete samples of local ULIRGs have found that the
great majority appear to be merging and/or disturbed systems (e.g., Murphy et
al. 1996; Clements et al. 1996). At lower IR luminosities (the LIRG category)
morphological studies have shown similarities with the ULIRG class although
the fraction of strongly interacting/disturbed systems appears to be smaller, approximately
$50-60\%$ (Wu et al. 1998). However, these LIRG studies have often been
conducted for incomplete samples or have been focused 
on the most luminous objects of this class (e.g., Wu et al. 1998;
Arribas et al. 2004).

The present study extends
the previous high-resolution near-IR morphological studies of LIRGs and ULIRGs (see Scoville et
al. 2001, and Bushouse et al. 2002) to a representative local sample where
most galaxies are at the lower luminosity end of the LIRG range. In our
sample, approximately 50\% of the
systems  are pairs, but only a small fraction are closely interacting/merging
galaxies (see Table~1). 

The most common continuum features in our sample are luminous 
star clusters and large-scale spiral arms.  
Two-armed spiral arm structures are also observed extending down to
the center (inner kpc scale, the so-called nuclear dusty spirals, see Martini
et al. 2003).
These spiral structures (both large and
small scale) are also present in at least one galaxy
member of close interacting systems or in
mergers (NGC~1614, IC~694, and IC~4687),
and pairs of  galaxies. In closely interacting systems a
more perturbed central morphology would be expected (and observed, see
Scoville et at 2000 and Bushouse et al. 2002), although this depends on
a number of factors, including relative sizes, and amount of gas present in the involved galaxies, 
and stage of the interaction.

In highly inclined galaxies in our sample dust lanes crossing the disks 
are common and present the reddest $m_{\rm F110W} - m_{\rm F160W}$ colors (see Fig.~1). 
If these dust lanes are hiding an old stellar
population (see Table~2), the typical extinctions would be in excess of $A_V=3-4\,$mag,
using the Rieke \& Lebofsky (1985) extinction law. 

The most ``extreme'' morphological features commonly observed in more IR
luminous systems (see Scoville et al. 2000 and Bushouse et al. 2002), such as
double/multiple nuclei systems with a large number of bright star clusters
near their centers and star-forming regions in the interface of interacting
galaxies, are not very common in our sample of LIRGs. The
most representative examples in this class are 
NGC~3690 (see Alonso-Herrero et al. 2000) and 
NGC~3256 (see Alonso-Herrero et al. 2002 and L\'{\i}pari et al. 2004). These
two systems 
are among the most IR luminous objects in our sample.

In terms of the central dust distribution, a few galaxies are dominated by nuclear
extinction: 
IC~860, MCG~$-$02-33-098E/W,
IC~4518W, and IRAS~17138$-$1017, whereas in other galaxies 
the regions of the largest extinction do not coincide with the nucleus of the
galaxy (e.g., NGC~3110, NGC~5734).
 In the cases of galaxies with nuclear  rings (see \S5.3) of star
formation, the H\,{\sc ii} regions
are interleaved with the dust features or the color maps trace a ring of
extinction just outside the ring of H\,{\sc ii} regions (e.g., NGC~23, and
NGC~1614, see Alonso-Herrero et al. 2001).


\begin{deluxetable*}{lcccc}

\tablewidth{8cm}

\footnotesize

\tablecaption{Observed 
photometry of the nuclei of LIRGs.}

\tablehead{Galaxy & $m_{\rm F160W}$ 
& $m_{\rm F110W}-m_{\rm F160W}$ &
$f_{\rm F187N}/f_{\rm F160W}$ & $
M_{\rm F160W}$  \\
(1) & (2) & (3) &(4) & (5) \\ 
\hline
\multicolumn{5}{c}{Nuclear point sources}}

\startdata
NGC~23           & 13.26 & 1.07 & 0.99 & $-20.6$\\
NGC~633          & 14.78 & 1.14 & 1.03 & $-19.4$\\
MCG~+12-02-001   & 13.61 & 1.37 & 1.15 & $-20.4$\\     
UGC~1845         & 12.41 & 1.48 & 1.21 & $-21.6$\\
NGC~2388         & 13.07 & 1.26 & 1.13 & $-20.7$\\
MCG~$-$02-33-098W & 14.26 & 1.22 & 1.19 & $-20.0$\\
IC~860           & 14.43 & 1.54 & 1.25 & $-19.6$\\
NGC~5135         & 14.46 & 1.45 & 1.54 & $-19.1$\\
NGC~5734         & 13.51 & 1.11 & 0.98 & $-20.4$\\
IC~4518W         & 14.40 & 1.95 & 1.74 & $-19.8$\\
IC~4686          & 13.29 & 0.92 & 1.01 & $-21.1$\\
NGC~6701         & 13.68 & 1.10 & 1.02 & $-20.1$\\
IC~5179          & 14.00 & 1.32 & 1.17 & $-19.4$\\
NGC~7771         & 14.99 & 1.67 & 1.32 & $-18.8$\\
\hline  
\multicolumn{5}{c}{Nuclear regions}\\
 \hline
UGC~3351         & 13.97 & 1.93 & 1.30 & $-20.0$\\
NGC~2369$^*$     & 14.23 & 1.59 & 1.30 & $-19.0$\\
MCG~+02-20-003   & 13.94 & 1.54 & 1.29 & $-20.2$ \\
NGC~3110         & 14.19 & 1.19 & 1.11 & $-20.1$\\
ESO~320-G030$^*$ & 13.62 & 1.28 & 1.11 & $-19.3$ \\
MCG~$-$02-33-098E & 14.64 & 1.30 & 1.19 & $-19.7$\\
IC~4518E         & 14.66 & 1.44 & 1.21 & $-19.6$\\
IRAS~17138$-$1017$^*$ & 14.74 & 1.87 & 1.38 & $-19.7$\\
IC~4687          & 14.24 & 1.33 & 1.14 & $-20.1$\\
IC~4734          & 13.27 & 1.43 & 1.19 & $-20.9$\\
NGC~7130         & 13.56 & 0.95 & 1.11 & $-20.5$\\
NGC~7591         & 13.49 & 1.43 & 1.21 & $-20.6$
\enddata
\tablecomments{
Column~(1): Galaxy name. Column~(2): Observed NICMOS F160W
  magnitude. Column~(3):  Observed $m_{\rm F110W}-m_{\rm F160W}$ color.
The typical color of an old
stellar population is $m_{\rm F110W}-m_{\rm F160W} = 1.03$,   
based on the integrated spectrum of an elliptical galaxy using the filter transmissions
and the quantum efficiency curves for the NICMOS filters.  
Column~(4): $1.87\,\mu$m to $1.60\,\mu$m continuum flux ratio. For the same
elliptical galaxy as in Column~(3), the $1.87\,\mu$m 
to $1.60\,\mu$m flux density ratio  is 
$f_{\rm F187N}/f_{\rm F160W} \simeq 0.85$. 
Column~(5): Absolute 
NICMOS F160W magnitude.\\ 
$^*$ Not obvious nucleus, the aperture was 
centered at the brightest
F187N source. In the case of IRAS~17138$-$1017 this corresponds to the
northern 
nucleus (see Zhou, Wynn-Williams, \& Sanders 1993).\\
In addition, near-IR 
nuclear point sources are present in NGC~7469 (Seyfert 1 nucleus: 
$M_{\rm F160W}=-22.9$, Scoville
et al. 2000),
NGC~1614 ($M_{\rm F160W}=-22.3$), NGC~3690 (B1: $M_{\rm F160W}=-19.5$ and
B2: $M_{\rm F160W}=-20.2$), NGC~5653 ($M_{\rm F160W}=-20.8$), 
and NGC~3256 ($M_{\rm F160W}=-20.2)$.} 
\end{deluxetable*}

\subsection{Nuclei of Galaxies}

Scoville et al. (2000) and Bushouse et al. (2002) 
found near-IR point-like nuclear sources in approximately $30-50$\%
of the galaxies in their samples of IR bright galaxies (i.e., some LIRGs but mostly 
ULIRGs). Scoville et al. (2000)
established that in about one-third of their sample these 
nuclear point sources dominate the near-IR emission at $2.2\,\mu$m. 
These bright nuclear point sources are more prevalent in 
LIRGs and ULIRGs with Seyfert or QSO activity. This is not surprising, as 
bright near-IR nuclear point  sources are common in nearby Seyfert 1 
galaxies, and are present in about 50\%
of nearby Seyfert 2 galaxies (see Quillen et al. 2001). However, galaxies with
no evidence of AGN activity can also show nuclear point sources, believed to
be nuclear star clusters. The typical luminosities of the nuclear star
clusters in late type spiral galaxies 
are in the range $10^6-10^7\,$L$_\odot$ (B\"oker et al. 2004 and
references therein), similar to those of the so-called super star clusters,
and more luminous galaxies tend to harbor more luminous nuclear clusters.

We have found that approximately 70\% of the 
galaxies in our
sample (including the galaxies observed prior to Cycle 13) 
show nuclear point sources, with observed (not corrected for
extinction) absolute $H$-band magnitudes ranging
from $M_{\rm F160W} = -22.9$ (for the bright Seyfert 1 nucleus in NGC~7469) 
to $M_{\rm F160W} = -18.8$, with an average value of 
$M_{\rm F160W}=-20.3$ (see Table~2). The observed
absolute $H$-band magnitudes for the resolved central regions (for sizes 
$170-370\,$pc, depending on the distance) are
similar to those of the nuclear point sources. In general, the observed  
range of F160W absolute magnitudes (Table~2) does not allow us to
discriminate between AGN  and star clusters. We also find that these point sources
do not dominate the near-IR emission except in some of the most IR luminous
galaxies (e.g., at $2.2\,\mu$m the Seyfert 1 nucleus of NGC~7469 see Scoville
et al. 2000, and component B1 in Arp~299 see Alonso-Herrero et al. 2000), or
in the case of the compact galaxy IC~4686.

The majority of the nuclei display 
redder colors (see Table~2) than those typical of an old stellar population, implying extinctions of
a few magnitudes (or higher if the stellar population is young, see \S7.2) 
for a simple dust screen model.
Scoville et al. (2000) found values of the nuclear $m_{\rm F110W} - m_{\rm
  F160W}$ colors  of between 1 and 2 for both LIRGs and ULIRGs in their
sample, thus similar to the colors observed in our sample. Scoville et
al. (2000) also found that the 
$m_{\rm F160W} - m_{\rm F222M}$ colors tend to be redder for the ULIRGs than
the less luminous LIRGs in their sample. We do not see any clear evidence for
this behavior, although our sample, unlike that of Scoville et al. (2000), 
does not contain examples of the most extreme cases of ULIRGs hosting QSOs,
known to present very red near-IR colors.

\begin{figure}



\includegraphics[angle=-90,width=8.5cm]{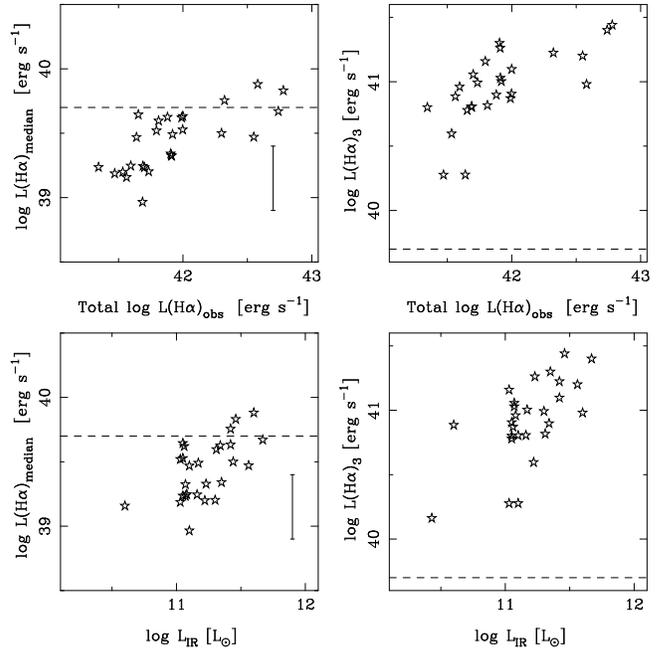}
\caption{{\it Left panels:} H$\alpha$ luminosity 
of the median H\,{\sc ii} region as a function of the IR luminosity (bottom)
and total observed (not corrected for extinction) H$\alpha$ luminosity of the
galaxy (top) in H\,{\sc ii} regions in the central regions. 
The horizontal dashed line shows the H$\alpha$
luminosity of 30 Doradus in the LMC, the prototypical giant H\,{\sc ii}
region. The error bar indicates the typical 
standard deviation from the mean value 
of the distribution of H\,{\sc ii} region H$\alpha$
luminosities of a given galaxy 
{\it Right panels:} Average H$\alpha$ luminosity 
of the three brightest  (first-ranked) H\,{\sc ii} regions as a function 
of the IR luminosity (bottom)
and total observed (not corrected for extinction) H$\alpha$ luminosity of the
galaxy (top). }
\end{figure}

\section{Morphology of the Pa$\alpha$ line emission of the central regions of LIRGs}

The {\it HST}/NICMOS  Pa$\alpha$ observations presented in this study afford
us the best opportunity to study  the H\,{\sc ii} region emission with spatial
resolutions of $25-50\,$pc in a sample of local LIRGs. 
The advantage of the Pa$\alpha$ observations when
compared  to optical H$\alpha$ observations is that the extinction 
is reduced by a
factor of approximately 5 ($A_{{\rm Pa}\alpha} \simeq 0.18 A_{{\rm H}\alpha}$, 
using the Rieke  \& Lebofsky 1985 extinction law).
In very dusty systems, such as those under consideration here, high spatial
resolution mid-IR imaging has shown the location (nuclear regions) and 
extent (approximately on scales of 1\,kpc or less) of the current dusty 
star formation 
(Soifer et al. 2001). It is of interest to determine if the Pa$\alpha$
emission 
is able to penetrate the high column densities traced by the
mid-IR emission.

We have grouped the main morphological Pa$\alpha$ features present
in the LIRG sample (see middle panels of Fig.~1) in five categories. 
 We note that some
galaxies show morphologies that place them in more than one category (see
Table~3). 

\subsection{Compact ($\lesssim 1\,$kpc) nuclear Pa$\alpha$ emission} 

The  Pa$\alpha$ morphology appears dominated by the nuclear emission (see
Table~3 for the galaxies in this class) with
sizes of less than approximately 1\,kpc, although
fainter H\,{\sc ii} regions are also present outside of the nuclear
regions. In some cases this compact emission appears to be related to the
presence of an AGN (see Table~1, and also \S9.2.2). 
The average observed H$\alpha$ surface brightnesses (derived from Pa$\alpha$) 
are among the highest in our sample of
galaxies: 
$5-10 \times 10^{41}\,{\rm erg \, s}^
{-1}$ kpc$^{-2}$. 

\subsection{Nuclear mini-spiral (inner $1-2\,$kpc) Pa$\alpha$ emission} 
In these cases we observe bright H\,{\sc ii} regions as well as 
bright star clusters along a  nuclear 
mini-spiral structure (see Table~3) 
also detected in continuum light (see \S4.1, and Martini et al. 2003). 
The typical observed  H$\alpha$ surface brightnesses are similar to the
  galaxies with compact Pa$\alpha$ emission.

\tablewidth{16cm}
\begin{deluxetable*}{lcccccc}

\footnotesize
\tablecaption{Statistics of the H\,{\sc ii } regions  
in the central regions of LIRGs.}

\tablehead{Galaxy 
& Pa$\alpha$ &  $\log L({\rm H}\alpha)_{\rm median}$ & 
$\log L({\rm H}\alpha)_{\rm br}$  & $\log L({\rm H}\alpha)_3$  & 
$L({\rm H}\alpha)$$_3$/$L({\rm H}\alpha)_{\rm HII}$ &
$L({\rm H}\alpha)_{\rm HII}$ \\
& morph. & erg s$^{-1}$ & erg s$^{-1}$ & erg s$^{-1}$ & 
& $\times 10^{40}$erg s$^{-1}$\\
(1) & (2) & (3) &(4) & (5) & (6) & (7) }
\startdata
NGC~23           & R &39.53 &40.94 &40.90 &  0.24 &  99.6 \\
MCG~+12-02-001   & M &39.50 &41.96 &41.60 &  0.61 & 199.4 \\
NGC~633          & R &39.16 &41.17 &40.88 &  0.63 &  36.5 \\
UGC~1845         & M &39.32 &41.23 &41.03 &  0.39 &  81.7 \\
UGC~3351         & E &39.20 &40.71 &40.60 &  0.35 &  34.1 \\
NGC~2369         & E &38.96 &40.87 &40.80 &  0.39 &  48.6 \\
NGC~2388         & E &39.33 &41.66 &41.26 &  0.67 &  81.4 \\
MCG~+02-20-003   & C &39.25 &41.31 &40.96 &  0.69 &  39.4 \\
NGC~3110         & S &39.60 &40.90 &40.82 &  0.30 &  64.8 \\
ESO~320-G030     & R &39.47 &40.31 &40.28 &  0.13 &  43.6 \\
MCG~$-$02-33-098 & C &\nodata &41.76 &\nodata & 0.95 & 101.6 \\
IC~860           & C &\nodata &40.13 &\nodata &  0.90 &   1.9 \\
NGC~5135         & M &39.49 &41.05 &41.00 &  0.36 &  83.1 \\
NGC~5734         & S &39.19 &40.52 &40.28 &  0.19 &  29.6 \\
IC~4518W         & C &39.52 &41.28 &41.16 &  0.69 &  62.4 \\
IC~4518E         & E &\nodata &40.21 &40.16 &  0.40 &  10.8 \\
NGC~5936         & M, S &39.23 &41.24 &41.05 &  0.68 &  50.3 \\
IRAS~17138$-$1017 & M &39.63 &41.24 &41.10 &  0.38 &  99.9 \\
IC~4687          & S &39.76 &41.42 &41.22 &  0.24 & 211.0 \\
IC~4686          & C &\nodata &41.63 & \nodata &  0.91 &  82.1 \\
IC~4734          & C, S &39.20 &41.02 &40.99 &  0.54 &  54.3 \\
NGC~6701         & M &39.64 &40.91 &40.78 &  0.40 &  45.1 \\
NGC~7130         & S &39.34 &41.69 &41.30 &  0.74 &  80.4 \\
IC~5179          & S &39.24 &41.18 &40.81 &  0.39 &  49.0 \\
NGC~7591         & C, S &39.24 &41.06 &40.80 &  0.85 &  22.1 \\
NGC~7771         & R, E, S &39.62 &41.05 &40.90 &  0.31 &  76.0 \\
\hline
NGC~1614         & R &39.88 &40.99 &40.98 &  0.08 & 380.2 \\
NGC~3256         & S &39.47 &41.49 &41.20 &  0.40 & 120.2 \\
NGC~3690         & S &39.83 &41.54 &41.44 &  0.44 & 186.2 \\
IC~694           & S &39.67 &41.82 &41.40 &  0.66 & 114.8 \\
NGC~5653         & S &39.62 &41.15 &40.87 &  0.23 &  97.7 \\
Zw~049.057       & E & \nodata & 41.32 & \nodata &  $\simeq 1$ &  20.9 

\enddata
\tablecomments{Column~(1): Galaxy Name. Column~(2): Pa$\alpha$
  morphology described in \S5 . C=compact emission. M=nuclear mini-spiral. R=ring of star formation. 
S=H\,{\sc ii} regions in spiral arms. E=edge-on galaxy.
Column~(3): H$\alpha$ luminosity of the median
H\,{\sc ii} region for galaxies with more than twenty H\,{\sc
  ii} regions detected. The typical standard deviations from the mean value of
the distribution
of H$\alpha$ luminosities of the H\,{\sc ii} regions detected is
$0.4-0.6\,$dex for a given galaxy. 
Column~(4): H$\alpha$ luminosity of the brightest
H\,{\sc ii} region. Column~(5): Average H$\alpha$ luminosity of the brightest three
(first-ranked)  H\,{\sc ii} regions. The typical uncertainties for the 
{\sc sextractor} fluxes for the brigthest and first-ranked H\,{\sc ii} regions
are $1-2\%$.
Column~(6): Ratio of the sum of the H$\alpha$ luminosities of 
the brightest three 
H\,{\sc ii} regions to the H$\alpha$ luminosity  in H\,{\sc ii} regions. 
Column~(7): Observed H$\alpha$ luminosity in H\,{\sc ii} regions in the
central regions. \\
The first part of the table are the galaxies observed in
Cycle 13, whereas the second part are the properties of the 
galaxies observed prior to Cycle 13 and presented
in Alonso-Herrero et al. (2002), recomputed for the distances used in this paper.}
\end{deluxetable*}

\subsection{Nuclear star-forming rings}
The nuclear rings in our sample of LIRGs (Table~3)
have  diameters ranging from $0.7\,$kpc to $2\,$kpc, 
show  resolved H\,{\sc ii} regions within
them, and dominate the central Pa$\alpha$ emission in
these galaxies. Fainter H\,{\sc ii} regions are also present in the
spiral arms in these systems. 
The typical observed (not corrected for extinction)  
H$\alpha$ surface brightnesses in the  H\,{\sc ii} regions 
are: $2-4 \times 10^{41}\,{\rm erg \, s}^
{-1}$ kpc$^{-2}$, although the ring of star formation in 
NGC~1614 has an H$\alpha$ surface brightness 
of $\simeq 60 \times 10^{41}\,{\rm erg \, s}^
{-1}$ kpc$^{-2}$. All these nuclear rings of star formation except 
the one in ESO~320-G030  are located in interacting galaxies. 

\begin{deluxetable*}{lccccc}
\footnotesize
\tablewidth{15cm}
\tablecaption{Pa$\alpha$ and H$\alpha$ fluxes, and extinctions.}
\tablehead{Galaxy & Aperture & $f({\rm H}\alpha)$ & 
$f({\rm Pa}\alpha$) & $A_V$ & $A_V$ (Veilleux)\\
& & erg cm$^{-2}$ s$^{-1}$ & erg cm$^{-2}$ s$^{-1}$ & mag & mag\\
(1) & (2) & (3) & (4) & (5) & (6)}
\startdata
NGC~23 & $2\arcsec \times 7\arcsec$ (E-W) & 
$2.5 \times 10^{-13}$ & $7.64 \times 10^{-14}$ & $1.7\pm0.4$ & 2.3\\
NGC~3110 & $2\arcsec \times 6\arcsec$ (E-W) & 
$^*7.2 \times 10^{-14}$ & $4.64 \times 10^{-14}$ & $3.1\pm0.4$ & 2.8\\
MCG~$-$02-33-098E & $2\arcsec \times 6\arcsec$ (N-S) & 
$^*3.0 \times 10^{-14}$ & $7.22 \times 10^{-14}$ & $5.5\pm0.4$ & 4.5\\
MCG~$-$02-33-098W & $2\arcsec \times 6\arcsec$ (N-S) & 
$^*4.4 \times 10^{-14}$ & $1.05 \times 10^{-13}$ & $5.5\pm0.4$ & 2.4\\
NGC~5734 & $2\arcsec \times 7\arcsec$ (E-W) & 
$2.8 \times 10^{-14}$ & $3.48 \times 10^{-14}$ & $4.3\pm0.4$ & 3.6\\
NGC~5936 & $2\arcsec \times 7\arcsec$ (E-W) & 
$8.9 \times 10^{-14}$ & $1.11 \times 10^{-13}$ & $4.3\pm0.4$ & 4.7\\
NGC~6701 & $1.5\arcsec \times 7\arcsec$ (E-W) & 
$1.0 \times 10^{-13}$ & $6.32 \times 10^{-14}$ & $3.1\pm0.4$ & 3.9\\
NGC~7130 & $1.5\arcsec \times 6\arcsec$ (E-W) & 
$2.1 \times 10^{-13}$ & $1.10 \times 10^{-13}$ & $2.7\pm0.4$ & 3.3\\
IC~5179 & $2\arcsec \times 9\arcsec$ (E-W) & 
$4.4 \times 10^{-14}$ & $8.69 \times 10^{-14}$ & $5.1\pm0.4$ & 3.7\\
NGC~7591 & $1.5\arcsec \times 6\arcsec$ (E-W) & 
$2.4 \times 10^{-14}$ & $4.42 \times 10^{-14}$ & $5.0\pm0.4$ & 4.3\\
NGC~7771 & $1.5\arcsec \times 7\arcsec$ (E-W) & 
$5.9 \times 10^{-14}$ & $8.28 \times 10^{-14}$ & $4.5\pm0.4$ & 6.3\\

\enddata
\tablecomments{Column~(1): Galaxy Name. 
Column~(2):  
Extraction aperture of the optical spectroscopy. 
Column~(3): Observed H$\alpha$ flux  from Veilleux et al.
(1995). ``$^*$'' next to the H$\alpha$ fluxes indicates that the optical 
spectra were obtained under
non-photometric conditions (Kim et al. 1995). Column~(4): Observed Pa$\alpha$ flux for the same 
aperture. Column~(5): Spectroscopic extinction derived from the H$\alpha$ and 
Pa$\alpha$ fluxes. The errors are computed for a 
15\% uncertainty in the Pa$\alpha$ emission line flux measurements. Column~(6):
Spectroscopic extinction derived from H$\alpha$ and H$\beta$ by Veilleux et
al.  (1995) for the same extraction apertures.}
\end{deluxetable*}

\subsection{Large-scale (several kpc) Pa$\alpha$ emission with H\,{\sc ii} regions
located in the spiral arms} 

In this category we find LIRGs with H\,{\sc ii} regions located over
several kpc along the spiral arms of the galaxy (see Table~3). 
Although our images only cover the central
$\simeq 3.3-7.2\,$kpc, several of these galaxies are known to show 
H$\alpha$ emission over larger scales (see
\S8, and Lehnert \& Heckman 1995;
Dopita et al. 2002; Hattori et al. 2004). 
The observed H$\alpha$ surface brightnesses in H\,{\sc ii} regions are about
one order of magnitude fainter than those of galaxies with compact Pa$\alpha$ emission.

In some cases the nuclear compact Pa$\alpha$ emission is bright and can make a
significant contribution to the total H\,{\sc ii} region emission, 
for instance in IC~694 and NGC~3690 
(see Alonso-Herrero et al. 2000) and NGC~3256 (see 
Alonso-Herrero et al. 2002). In this class we also find examples of galaxies
without bright nuclear emission and large numbers of bright H\,{\sc ii}
regions in the spiral arms (IC~4687), and even galaxies whose Pa$\alpha$ emission
is dominated by one or a few 
bright extra-nuclear H\,{\sc ii} regions: NGC~5653 (see B\"oker et al. 1999 and
Alonso-Herrero et al. 2002) and NGC~5734.
 
\subsection{Large-scale Pa$\alpha$ emission in highly inclined dusty systems} 
In these cases (see Table~3)
the Pa$\alpha$ emission traces the least
obscured H\,{\sc ii} regions along the disk 
of the galaxies over scales of a few kpc (3.3 to 7.2\,kpc, 
at least, see \S8). This category is similar to that discussed in \S5.4 but
for galaxies viewed edge-on.
These galaxies tend to show the reddest $m_{\rm
  F110W}-m_{\rm F160W}$ central colors
over the extent of the Pa$\alpha$ emission (see right panels of Fig.~1).

Summarizing, over half the galaxies in our sample 
most of the Pa$\alpha$ emission is more compact than the 
continuum near-IR  emission, and 
is located in the inner $1-2\,$kpc of the galaxies, as also shown by
mid-IR observations (Soifer et al. 2001) of a few bright LIRGs. 
In a few cases the nuclear emission even dominates the total Pa$\alpha$
emission (see further discussion in \S8). However, 
in the other half of  our sample the
Pa$\alpha$ emission is extended on scales larger than a few kpc (as
discussed in \S5.4 and \S5.5).

\section{Properties of the H\,{\sc ii} regions in LIRGs}

Alonso-Herrero et al. (2002) showed for a sample of seven\footnote{Two
  galaxies in their sample, VV~114 and NGC~6240, are at $v>5200\,{\rm km
  \,s}^{-1}$, and thus are not included in our sample.} LIRGs that these galaxies contained a
significant number of exceptionally bright H\,{\sc ii} regions (based on the
median values of the  
H\,{\sc ii} region luminosity distribution) with H$\alpha$ luminosities
comparable to that of
the  giant H\,{\sc ii}  region 30 Doradus ($\log L({\rm H}\alpha) = 39.70\,$[erg s$^{-1}]$,
Kennicutt, Edgar, \& Hodge 1989). This small sample was, however, heavily
biased towards the most IR luminous objects in the LIRG class. 
Table~3 summarizes the statistical properties of the H\,{\sc ii} regions
detected in our sample of LIRGs. 

Confirming the results of Alonso-Herrero et al. (2002), the LIRGs in our sample 
with the highest IR and observed H$\alpha$ luminosities have
median H\,{\sc ii} regions with H$\alpha$ luminosities comparable or brighter than the giant
H\,{\sc ii} region 30 Dor (Table~3, and left panels of Fig.~2). 
Approximately 50\% of our volume-limited sample of LIRGs
have $\log L({\rm H}\alpha)_{\rm median} > 39.5\,$[erg s$^{-1}]$, significantly higher
(about one order of magnitude) than the median H\,{\sc ii} regions of 
normal galaxies in the Virgo Cluster observed with  similar
spatial resolutions (see Alonso-Herrero \&
Knapen 2001).  

The first-ranked (the brightest three) H\,{\sc ii} regions are usually coincident with
the nuclei of the galaxies, although with notable exceptions, e.g., NGC~5653,
  NGC~5734, and IC~4687, as well as those galaxies with nuclear rings of star formation (see
  \S5.3). The average H$\alpha$ luminosities of the first-ranked H\,{\sc
  ii} regions are between 1 and 2 orders of magnitude brighter than the
median H\,{\sc ii} regions, and are well above that of 30 Dor (Table~3). 
The contribution of the first-ranked H\,{\sc ii} regions to the total observed H$\alpha$ luminosity in
H\,{\sc ii} regions in the central regions is given in Table~3. 
Not surprisingly, this contribution is
highest for the galaxies with compact Pa$\alpha$ emission where these reions
also tend to coincide with or are near the nucleus of the galaxy.

There is a good correlation between the H$\alpha$ luminosity of the
  median H\,{\sc ii} regions and the IR and total observed H$\alpha$
  luminosity of the system (left panels of Fig.~2).
These correlations appear to be linear. We also show in these figures 
the typical standard deviation from the
mean value of the distribution of H$\alpha$ luminosities of the detected H\,{\sc
  ii} regions for a given galaxy, an indication of the width of the distribution. 
Similarly, there is a good relation between
the luminosity of the first-ranked regions 
 and the IR and the total 
H$\alpha$ luminosity in H\,{\sc ii} regions (right panels of Fig.~2). The uncertainties for
the {\sc sextractor} H$\alpha$ fluxes  of the first-ranked H\,{\sc ii} 
regions ($1-2\%$) plus those of the 
distance determinations ($\sim 10\%$) 
of the galaxies do not account for the dispersion in
H$\alpha$ luminosity of the first-ranked (and also median) 
H\,{\sc ii} regions for a given IR luminosity. This dispersion 
is probably  due to resolution effects, as
the distances of our galaxies span a factor of two.

\begin{deluxetable*}{lcccc}

\footnotesize
\tablewidth{15cm}
\tablecaption{Pa$\alpha$ and Br$\gamma$ fluxes, and extinctions.}
\tablehead{Galaxy & Aperture & $f({\rm Br}\gamma)$ & 
$f({\rm Pa}\alpha$) & $A_V$\\
& & erg cm$^{-2}$ s$^{-1}$ & erg cm$^{-2}$ s$^{-1}$ & mag\\
(1) & (2) & (3) & (4) & (5) }
\startdata
NGC~2388 & $3\arcsec \times 9\arcsec$ (E-W) & 
$26.8 \times 10^{-15}$ & $2.02 \times 10^{-13}$ & $13\pm5$\\
NGC~3110 & $3\arcsec \times 9\arcsec$ (E-W) &
$6.0 \times 10^{-15}$ & $0.68 \times 10^{-13}$ & $2\pm5$\\
NGC~3256-nucleus & $3.5\arcsec \times 3.5\arcsec$ &
$54.0\times 10^{-15}$ & $5.37\times 10^{-13}$ & $6\pm3$\\
NGC~3256-5\arcsec \, S & $3.5\arcsec \times 3.5\arcsec$ &
$15.0\times 10^{-15}$ & $1.15\times 10^{-13}$ & $15\pm5$\\
NGC~3256-5\arcsec \, E &  $3.5\arcsec \times 3.5\arcsec$& 
$17.0\times 10^{-15}$ & \nodata & \nodata\\
NGC~5135 & $3\arcsec \times 12\arcsec$ (E-W) &
$16.5 \times 10^{-15}$ & $1.95 \times 10^{-13}$ & $0.5\pm5$\\
IRAS~17138$-$1017 & $3\arcsec \times 9\arcsec$ (N-S) &
$22.4 \times 10^{-15}$ & $1.62 \times 10^{-13}$ & $14\pm5$\\
NGC~7130 & $1.5\arcsec \times 4.5\arcsec$ (E-W) &
$10.2 \times 10^{-15}$ & $1.00 \times 10^{-13}$ & $7\pm5$\\

\enddata
\tablecomments{Column~(1): Galaxy Name. The two other galaxies in common with
  Goldader et al. (1997a) are NGC~1614 analyzed in Alonso-Herrero et
  al. (2001), and NGC~7469 for which the spectroscopic data includes the Seyfert 1
  nucleus. 
Column~(2):  
Extraction aperture. Column~(3): Observed Br$\gamma$ flux  from Goldader et
al. (1997a) for all galaxies except for NGC~3256 for which the data are from
Doyon et al. (1994). For NGC~3256 the 5\arcsec \ to the south location coincides with the secondary
nucleus of the galaxy, and the 5\arcsec \ to the east location, an H\,{\sc ii} region, is
not fully covered by our Pa$\alpha$ image. 
Column~(4): Observed Pa$\alpha$ flux for the same 
aperture. Column~(5): Derived spectroscopic extinction. The errors are computed for a 
15\% uncertainty in the Pa$\alpha$ emission line flux measurements.}
\end{deluxetable*}

\begin{deluxetable}{lccc}

\footnotesize

\tablecaption{Photometric extinctions, extinction-corrected
  Pa$\alpha$ luminosities, and SFRs.}
\tablehead{Galaxy & 
Phot $A_V$ &  $\log L({\rm Pa}\alpha)_{\rm corr}$&  
$\log$ SFR ${\rm IR}$/N$_{\rm Ly}$ \\
       & mag & erg s$^{-1}$ \\
(1) & (2) & (3) & (4)}
\startdata
NGC~23           &$2.7^{+  1.1}_{-  0.8}$& 41.32 & 0.14 ($-$0.01)\\
MCG~+12-02-001      &$4.5^{+  1.0}_{-  0.8}$& 41.67 & 0.17\\
NGC~633          &$2.6^{+  1.0}_{-  0.8}$& 40.91 & 0.10\\
UGC~1845         &$4.6^{+  1.1}_{-  0.8}$& 41.33 & 0.14\\
UGC~3351         &$5.9^{+  1.0}_{-  0.8}$& 41.25 & 0.38 (0.23)\\
NGC~2369         &$4.8^{+  1.1}_{-  0.9}$& 41.16 & 0.34\\
NGC~2388         &$3.5^{+  1.1}_{-  0.8}$& 41.29 & 0.35\\
MCG+02-20-003      &$3.3^{+  1.0}_{-  0.8}$& 41.16 & 0.33\\
NGC~3110         &$3.3^{+  1.1}_{-  0.8}$& 41.26 & 0.45 (0.15)\\
ESO~320-G030     &$3.3^{+  1.1}_{-  0.8}$& 40.97 & 0.53\\
MCG$-$02-33-098      &$3.3^{+  1.1}_{-  0.8}$& 41.32 & 0.20\\
IC~860           &$3.2^{+  0.8}_{-  0.7}$& 39.86 & 1.72\\
NGC~5135         &$3.3^{+  1.1}_{-  0.9}$& 41.26 & 0.32\\
NGC~5734         &$3.1^{+  1.2}_{-  0.9}$& 41.05 & 0.39\\
IC~4518W         &$4.1^{+  0.8}_{-  0.6}$& 41.27 & 0.17\\
IC~4518E         &$4.6^{+  1.2}_{-  0.9}$& 40.72 & 0.12\\
NGC~5936         &$3.7^{+  1.3}_{-  0.9}$& 41.20 & 0.27\\
IRAS~17138$-$1017  &$5.1^{+  0.7}_{-  0.8}$& 41.48 & 0.34\\
IC~4687          &$3.1^{+  1.2}_{-  0.9}$& 41.63 & 0.19\\
IC~4686          &$2.2^{+  1.0}_{-  0.7}$& 41.21 & 0.15\\
IC~4734          &$3.9^{+  1.2}_{-  0.9}$& 41.19 & 0.52\\
NGC~6701         &$3.0^{+  1.1}_{-  0.9}$& 41.04 & 0.42 (0.27)\\
NGC~7130         &$2.9^{+  1.2}_{-  0.9}$& 41.30 & 0.45\\
IC~5179          &$3.3^{+  1.2}_{-  0.8}$& 41.22 & 0.35 (0.13)\\
NGC~7591         &$4.1^{+  1.2}_{-  0.9}$& 41.09 & 0.36\\
NGC~7771         &$4.1^{+  1.2}_{-  0.9}$& 41.36 & 0.39 (0.04)\\
\enddata

\tablecomments{Column~(1): Galaxy name. Column~(2): Average photometric extinction 
$A_V$ over the Pa$\alpha$ emitting region derived from the
  observed $m_{\rm F110W}-m_{\rm F160W}$ colors (see \S7.2).
The uncertainties of phot $A_V$ are for the possible range of ages of the Pa$\alpha$ emitting
  regions (4\,Myr and 9\,Myr) using the Rieke et al. (1993) models for a
  Gaussian burst with FWHM=5\,Myr (see Alonso-Herrero et al. 2002 for
  details).
 We also list the Pa$\alpha$ luminosities
over the NICMOS FOV corrected for extinction. and uncertainties. 
Column~(3) Pa$\alpha$ luminosity corrected for extinction. 
Column~(4): Log of the ratio between the
IR and corrected for extinction 
N$_{\rm Ly}$ SFRs calculated using the prescriptions of Kennicutt (1998). 
In brackets we give the values corrected for extended
emission (\S8, and Table~7).}

\end{deluxetable}

\begin{deluxetable}{lccc}


\tablewidth{8cm}

\footnotesize
\tablecaption{Large-scale emission.}
\tablehead{Galaxy & 
Diameter  & \multicolumn{2}{c}{$\frac{f_{\rm NIC2\,FOV}}{f_{\rm tot}}$}\\  
          & H$\alpha$ & H$\alpha$ & $24\,\mu$m\\
(1) & (2) & (3)}
\startdata
NGC~23           & 55 & 0.70 & 0.83 \\
UGC~3351         & \nodata & \nodata & 0.73\\
NGC~3110         & 76 & 0.50 & 0.63\\
NGC~6701         & \nodata & \nodata & 0.73 \\   
IC~5179          & \nodata & \nodata & 0.61 \\
NGC~7771 & 93 & 0.44 & 0.55\\
\enddata
\tablecomments{Column~(1): Galaxy. 
Column~(2): Diameter in arcsec used to measure the
  total H$\alpha$ emission. Column~(3): Ratio of the observed H$\alpha$
  emission over the NICMOS NIC2 FOV ($\sim 
20\arcsec \times 20\arcsec$) and the total observed H$\alpha$ emission 
using the images of Hattori et al. (2004). Column~(4): Ratio of the {\it
  Spitzer}/MIPS $24\,\mu$m 
emission over the NICMOS FOV and the entire galaxy.} 
\end{deluxetable}

\section{Extinction Correction of Pa$\alpha$ measurements}

We now prepare to compare the star formation rates (SFR) derived
from the IR luminosity and from the number of ionizing photons (N$_{\rm Ly}$)
 (see \S9). To do so, we must correct the
observed hydrogen recombination line emission for extinction (see discussion
in Kennicutt 1998). Since the mid-IR  emission of (some) LIRGs 
is predominantly produced in nuclear starbursts with sizes of less than
approximately $1-2\,$kpc (see e.g., Wynn-Williams \& Becklin 1993; 
Miles et al. 1996; Keto et al. 1997; Soifer et al. 2001), there must be 
large concentrations of dust in their  nuclear regions. 

 The largest uncertainty in determining the extinction by far comes from the unknown 
distribution of dust within the source. The color maps
(Fig.~1) of most LIRGs in our sample show complicated dust features in their
centers. In the cases of the deeply embedded star-forming regions, common in
LIRGs and ULIRGs, a simple foreground dust screen model may not provide a
good fit to the star-forming properties (e.g., see Genzel et al. 1995;
Genzel et al. 1998; 
Satyapal et al. 1999), and a model in which stars and dust are mixed will be
more realistic (Witt \& Gordon 2000).
 Moreover, the simple model of a foreground dust screen 
only provides a  lower limit to
the true obscuration, and in cases of high optical depths both models will
only measure the obscuration to the material less affected by extinction
(Goldader et al. 1997a). Because of this, in very dusty systems there is a tendency for  
the derived extinction to increase as longer wavelengths are used (see e.g.,
McLeod et al. 1993; Genzel et al. 1998). Finally there is the issue of 
whether the extinction to the stars is the same as the extinction to
the gas. Calzetti, Kinney, 
\& Storchi-Bergmann (1994) found that in general the extinction to
the gas  (ionized by the young stars) is twice that of the continuum (mostly
produced by old stars), implying  that the youngest 
ionizing stars are located in dustier regions than the cold stellar population is.
The lack of correspondence between gas emission and continuum emission
generally observed in LIRGs (Fig.~1) suggests that this may also be the case for them.

\subsection{Extinction to the gas}

Kim et al. (1995) and Veilleux et al. (1995), and 
Goldader et al. (1997a,b) presented the most comprehensive optical and near-IR 
long-slit spectroscopic
surveys, respectively, of the central regions of LIRGs. 
We have a number of galaxies in common with these studies for which we can
estimate the extinction to the gas. 
We have simulated the sizes and orientations of the slits used by Kim et al. (1995) and Goldader et
al. (1997a) to measure the Pa$\alpha$ fluxes and compared them with
their H$\alpha$ and Br$\gamma$ measurements, respectively. 
We have then estimated the extinction to the gas using the
Rieke \& Lebofsky (1985) extinction law and a foreground dust screen model.  
In addition, for NGC~3256 we have used the
Br$\gamma$ fluxes of Doyon et al. (1994).

The results are presented in Tables~4 and 5. Table~4 additionally lists
the $A_V$ derived from H$\beta$
and H$\alpha$ by Veilleux et al. (1995), which are generally similar 
to or slightly smaller than the $A_V$
obtained from H$\alpha$ and Pa$\alpha$. The extinctions derived 
using the longest wavelength emission lines (Table~5) tend to be higher for
the few galaxies in common with H$\alpha$ and Pa$\alpha$ measurements.
The average extinctions obtained from H$\alpha$ and Pa$\alpha$ over regions
with typical sizes of $0.5\,{\rm kpc} \times 1.5\,{\rm kpc}$ are not too
different from the $A_V$ (derived using the same emission lines) 
of  individual H\,{\sc ii} regions in spiral galaxies
and starburst galaxies (see e.g., Quillen \& Yukita 2001; Maoz et al. 2001;
Scoville et al. 2001; Calzetti et al. 2005). 


\subsection{Extinction to the stars}
We can alternatively use the observed near-IR colors to estimate 
the extinction to the young stars
responsible for ionizing the H\,{\sc ii} regions. The near-IR colors
in general present a smaller dependence on the age of the stellar population than the
optical colors. However, there is a weak dependence, 
in particular for the youngest stellar populations responsible for
ionizing the H\,{\sc ii} regions. Alonso-Herrero et al. (2002) found, for a
small sample of LIRGs, that most
young star-forming regions (with Pa$\alpha$ emission) 
in LIRGs have ages of $\simeq 1-9\,$Myr (depending
on the assumed type of star formation).


Using the typical near-IR colors of a young ionizing stellar population predicted by
the Rieke et al. (1993) models, and the observed  
F110W-F160W color maps (right panels of Fig.~1), we have 
derived 2-D extinction (refered to as photometric $A_V$) maps. The uncertainties 
of the extinction maps are given by the range of possible ages of the Pa$\alpha$ emitting regions.
The photometric $A_V$ maps
are only applicable to continuum regions 
with Pa$\alpha$ flux emission. For each galaxy, the extinction-corrected Pa$\alpha$
maps were constructed using the $A_V$ maps and the observed Pa$\alpha$ images. To
estimate the observed and extinction-corrected Pa$\alpha$ fluxes,  we only 
added those
pixels above the threshold used for estimating the fluxes of H\,{\sc ii}
regions in \S6. An estimate of the background is obtained from those pixels
below the threshold value. We note that in computing the observed Pa$\alpha$
fluxes in this fashion we are including more pixels than when measuring
H\,{\sc ii} region fluxes with SE{\sc xtractor} (\S3.3), as we are not imposing a criterion
for defining the size of the emitting regions. The typical uncertainties for
the summed Pa$\alpha$ fluxes are $\simeq 30\%$, depending on the
threshold value used for adding up the emission. 
The average photometric extinctions over the Pa$\alpha$ emitting regions
for our sample of galaxies are given in the second column of Table~6.

A comparison with the spectroscopic extinctions in Tables~4 and 5 shows that the
photometric extinctions are comparable to or slightly smaller than the extinctions to the
gas for the galaxies in common. However, in most cases, the photometric
extinctions are derived for larger areas than the regions covered by the
spectroscopic observations. Since there is evidence in spiral galaxies 
that the extinction tends
to decrease with distance from the galaxy center (Calzetti et al. 2005), the
average photometric extinction measured over larger regions will tend to be
smaller
than the spectroscopic values. We also find that the highest values of the
average photometric extinction over the Pa$\alpha$ emitting regions tend occur in those
galaxies with the most compact ($\simeq 1\,$kpc) Pa$\alpha$
emitting regions (e.g., MCG~+12-02-001) or in the edge-on
systems (e.g., NGC~2369, UGC~3351).

\section{Large-scale Emission}
As we saw in \S5 about half of the galaxies in our sample show extended
Pa$\alpha$ emission covering the entire FOV of the NICMOS observations. 
Before we compare the Pa$\alpha$ (or H$\alpha$) and IR luminosities, we need to assess the
importance of the H\,{\sc ii} region and diffuse emission at large galactocentric
distances, especially for the nearest
examples in our sample. 
Hattori et al. (2004) for their sample of $60\,\mu$m
flux-limited LIRGs, which includes some of the galaxies in our sample, 
have found that the star formation activity is dominated by
emission extending over several kpc. 
We have obtained the H$\alpha$ images from Hattori et
al. (2004) for the galaxies in
common with our work, and measured the H$\alpha$ fluxes within the NICMOS FOV and
the total extent of the galaxy. As can be seen from Table~7, the contribution from the
large-scale H$\alpha$ emission can be as high as 50\% in some galaxies. We have eight further galaxies in 
common with the sample of Dopita et al. (2002). From inspection of their H$\alpha$
images, which comprise the entire systems, we infer that the NICMOS Pa$\alpha$
images include most of the emission of NGC~633, 
MCG$-$02-33-098E/W, IC~4518E/W, NGC~5734, and IC~4686/IC~4687. 


The H$\alpha$ images of IC~4734, and 
NGC~7130 (Dopita et al. 2002), as well as  of IC~5179 (see 
Lehnert \& Heckman 1995) and NGC~6701 (see M\'arquez, Moles, \& Masegosa 1996)
show H\,{\sc ii} region and diffuse H$\alpha$ 
emission over scales larger than the FOV of the Pa$\alpha$ images, and thus our
extinction-corrected Pa$\alpha$ luminosities are lower limits to the total
emission. If the
average extinction of H\,{\sc ii} regions decreases for increasing
galactocentric distances (Calzetti et al. 2005), as in spiral galaxies,  
the contributions from the large-scale component not covered by the NICMOS images 
derived from the H$\alpha$ images will be upper limits.

We have retrieved the available MIPS $24\,\mu$m images from the {\it Spitzer} archive and
simulated the NICMOS FOV to estimate the contribution from the large-scale
emission. At $24\,\mu$m the NICMOS FOV ($\sim 20\arcsec \times 20\arcsec$) 
covers most of the emission
($>85\%$) for the majority of the galaxies of the sample, except for those
listed in Table~7. For those galaxies with available
H$\alpha$ imaging, the agreement on the fraction of extended emission derived
from the  $24\,\mu$m imaging is reasonably good. We conclude that, with a few
exceptions, the Pa$\alpha$ detected by NICMOS is representative of the totals
for our sample of galaxies.

\begin{figure*}
\includegraphics[angle=-90,width=16cm]{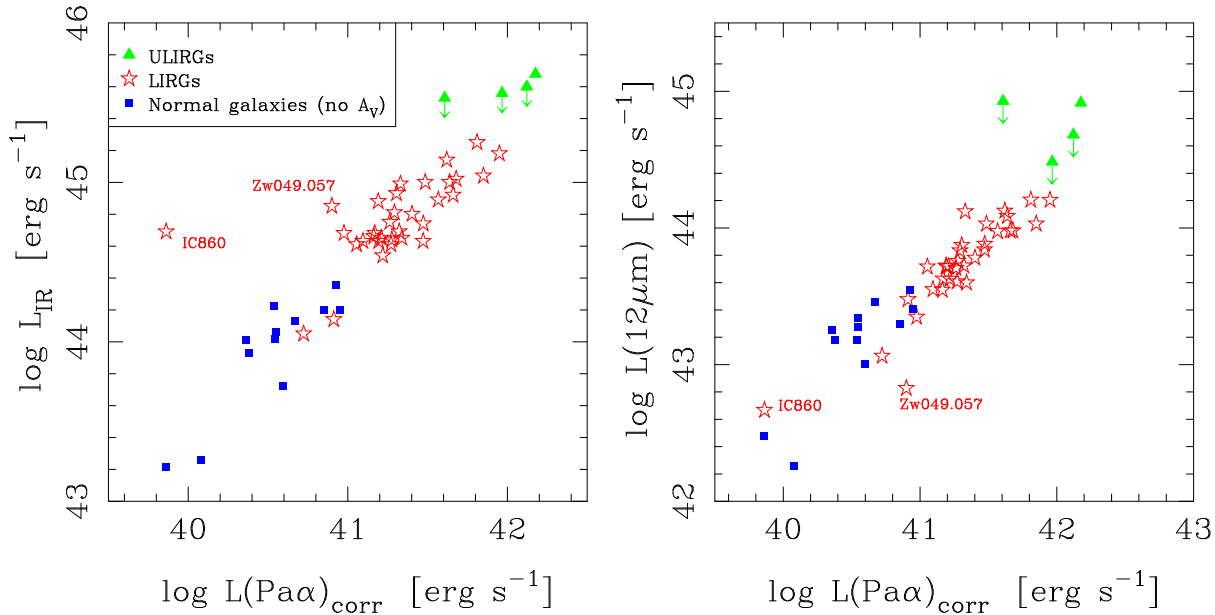}
\caption{{\it Left panel:} Comparison between the IR and the
  extinction-corrected 
Pa$\alpha$ luminosities. For NGC~7469 we have derived the
  extinction-corrected Pa$\alpha$ luminosities of the star-forming ring 
from the Br$\gamma$ imaging data of  Genzel et al. (1995). Note that three ULIRGs
  of the Murphy et al. (2001) sample are not detected at $12\,\mu$m, and thus
  their L$_{\rm IR}$ are shown as upper limits.
{\it Right panel:} Comparison between the monochromatic $12\,\mu$m and the extinction
  corrected Pa$\alpha$ luminosities. }
\end{figure*}

\section{Star Formation Rates of LIRGs}

Both the IR emission and $N_{\rm Ly}$ are good tracers of
the massive SFR in galaxies. $N_{\rm Ly}$ is usually
derived from hydrogen recombination lines, and has the advantage that it 
traces the current massive SFR, almost independently of the
previous star formation history of the galaxy (Kennicutt 1998). The main
disadvantage in deriving SFRs from $N_{\rm Ly}$ 
is determining $A_V$ and the fraction of escaping photons
from H\,{\sc ii} regions. The former problem can be alleviated by using near-IR,
mid-IR, and radio recombination lines (e.g., H92$\alpha$,  see 
Zhao et al. 1997; Roy et al. 2005). The issue of the 
fraction of escaping photons is still uncertain and
is not considered here, but it has  been addressed both observationally (Beckman et
al. 2000) 
and theoretically (Bland-Hawthorn \& Maloney 1999). 

The IR luminosity of a starburst galaxy is due
to UV emission (produced mainly by the young stellar population) 
absorbed by dust and re-emitted in the thermal IR. As discussed
by Kennicutt (1998), the efficacy of the IR luminosity as a tracer of the SFR
depends on the contribution of the young stars to the heating of the dust, and
requires that all the UV light from massive stars be absorbed by the dust
(i.e., $A_V>1\,$mag, see also Calzetti et al. 2005). In
the presence of high dust opacities, as in LIRGs, the IR luminosity dominates the bolometric
luminosity of the system and is the ultimate tracer of the SFR, but with a
somewhat ill-determined time scale.

\subsection{Comparison between the mid-IR  
and IR luminosities and the
  extinction-corrected Pa$\alpha$ luminosities}

The mid-IR luminosities are routinely used to estimate the total IR
luminosities and SFRs of galaxies  at cosmological distances (e.g., Elbaz et
al. 2002; P\'erez-Gonz\'alez
et al. 2005; Le Floc'h et al. 2005). Observationally the 
$12\,\mu$m luminosity is found to be a good indicator
of the total IR luminosity of local galaxies (e.g., Rush, Malkan, \& 
Spinoglio 1993; Elbaz et al. 2002; Takeuchi et al. 2005). The advantage in
using mid-IR luminosities as indicators of the SFR is that they are not
affected by the contribution from cold dust heated by old stars that
  may dominate the far-IR luminosities. {\it Spitzer} observations of nearby {\it normal} 
galaxies  are now showing  that there is a good correlation
between the Pa$\alpha$ or H$\alpha$ luminosity (corrected for extinction) and the
$24\,\mu$m luminosity of H\,{\sc ii} knots and H\,{\sc ii} 
regions (e.g., M51 Calzetti et al. 2005;
  M81 P\'erez-Gonz\'alez et al. 2006), indicating that the latter
luminosity could also be a good potential SFR tracer. Calzetti et
  al. (2005) noted,
 however, that the $24\,\mu$m luminosity to SFR ratios of UV-selected
  starbursts and ULIRGs deviate from the average value found for the inner
  region of M51 suggesting that differences among different types of galaxies
  are likely to be present.

Fig.~3 shows a comparison between the IR and the monochromatic {\it IRAS} $12\,\mu$m 
luminosities and the extinction-corrected Pa$\alpha$ luminosities for our
sample of LIRGs. The 12$\mu$m luminosities for close interacting systems are
based on Surace et al. (2004). When they could not obtain {\it IRAS} fluxes for
the individual components of close interacting systems, the $12\,\mu$m and IR
luminosities of each component were assumed to have a ratio similar to that 
of their Pa$\alpha$ luminosities.

In addition to our sample of LIRGs, we have compiled a small comparison sample of normal galaxies with
available NICMOS Pa$\alpha$ imaging observations (not corrected for
  extinction) from B\"oker et al. (1999). The normal galaxies were chosen so
  that the NICMOS Pa$\alpha$ observations covered the majority of the ionized hydrogen
emission.  If the values of the extinction inferred from H$\alpha$/Pa$\alpha$
line ratios for a few spiral galaxies
and star-forming galaxies ($A_V\sim 1-4\,$mag, 
see e.g., Maoz et al. 2001; Quillen \& Yukita 2001; Calzetti et al. 2005) are
  representative of our sample of normal galaxies, the extinction 
correction for the Pa$\alpha$
luminosity would only be up to $0.2\,$dex. Finally to extend the luminosity range
  we have included in this comparison the four ULIRGs imaged in Pa$\alpha$ by 
Murphy et al. (2001). For the ULIRGs the Pa$\alpha$ luminosities have been
  corrected for extinction using the $A_V$ values these authors derived
 from the H$\alpha$/Pa$\alpha$ line ratios.

There is a tight correlation between
both the IR and $12\,\mu$m, and the extinction-corrected Pa$\alpha$
luminosities, although some of the galaxies with deviating IR luminosities
(marked in the figures),
seem to follow better the correlation when the $12\,\mu$m luminosity is used. 
For reference we give in the last column of Table~6 
the ratio between the SFRs
derived from the IR luminosity and the extinction-corrected Pa$\alpha$
luminosity using the prescriptions of Kennicutt (1998). As can be seen from
this table, for local LIRGs 
the SFRs derived from the number of ionizing photons are on average
$0.2-0.3\,$dex lower than those inferred from the total IR luminosity. This
behavior is consistent with the tendency for the measured IR luminosity to
include some contribution from older stars (see below).

\subsection{The $24\,\mu$m emission as a SFR tracer for dusty star-forming galaxies}
We now examine in more detail the relation between mid-IR $24\,\mu$m and Pa$\alpha$
luminosities. The monochromatic {\it IRAS} $25\,\mu$m 
luminosities  for our sample of LIRGs, normal galaxies, and ULIRGs have been
converted to monochromatic $24\,\mu$m luminosities using the Dale \& Helou
(2002) models with appropriate indices for the dust distribution (in their
notation we use $\alpha=1.5$ and $\alpha=2$ for LIRGs and normal galaxies,
respectively). In addition we use {\it Spitzer}/MIPS $24\,\mu$m observations of resolved star-forming regions
within the central 6\,kpc region of M51 from Calzetti et
al. (2005), and for M81 (entire galaxy) from P\'erez-Gonz\'alez et al. (2006).
For M51, Calzetti et
al. (2005) have derived the extinction corrections from H$\alpha$/Pa$\alpha$
line ratios, and for M81 
we transform the extinction-corrected (based on $A_V$ derived from the Balmer decrement
or radio) H$\alpha$ luminosities (see
P\'erez-Gonz\'alez et al. 2006
for more details) to Pa$\alpha$ assuming Case B recombination. The inclusion
of data for resolved star-forming regions within M51 and M81 allows us to extend
the range of the mid-IR-Pa$\alpha$ relation down three  orders  of magnitude.

As can be seen from Fig.~4 (left panel), the LIRGs, ULIRGs, and normal galaxies
seem to continue the relation between the extinction-corrected Pa$\alpha$
luminosity and the $24\,\mu$m luminosity observed for the M51 central H\,{\sc ii}
knots. We also show in this figure data for
galaxies from the
Universidad Complutense de Madrid sample (UCM, P\'erez-Gonz\'alez et al. 2003) and 
the Nearby Field Galaxy Sample (NFGS, Kewley et al. 2002)
 with available H$\alpha$ imaging, and extinction corrections derived
from the Balmer decrement. These galaxies also follow the trend but with a
larger scatter, as their measurements are from H$\alpha$ imaging rather than
Pa$\alpha$. The M81 H\,{\sc ii} regions, on the other hand,
follow a similar linear relation,
but the  relation appears to be offset with respect to that of M51, normal
galaxies,  and LIRGs
 (see discussion by P\'erez-Gonz\'alez et al. 2006). This
behavior could arise from a lower UV absorption efficiency in the relatively
low luminosity and lightly obscured H\,{\sc ii} regions in M81.

The best fit to the extinction-corrected Pa$\alpha$ vs. $24\,\mu$m luminosity
relation for resolved star-forming regions of
M51, normal galaxies (with available Pa$\alpha$ imaging), 
LIRGs (excluding IC~860), and ULIRGs is:

\begin{equation}
  \log L(24\mu{\rm m}) \\
= (-3.553\pm0.516)+(1.148\pm0.013) \times \log L({\rm
  Pa}\alpha)_{\rm corr}
 \end{equation}

\noindent where the luminosities are in erg s$^{-1}$, and the $24\,\mu$m luminosity is
computed as the monochromatic value, i.e., from $\nu f_\nu$.  
For comparison the fit to the M51 H\,{\sc ii} regions alone gives a slightly smaller
slope ($1.088\pm0.061$, see also Calzetti et al. 2005), 
but within the errors of the  fit including normal
galaxies, LIRGs, and ULIRGs. The fit to the M81 H\,{\sc ii} regions
alone provides a similar slope, 
essentially equal to unity ($0.987\pm0.064$).

\begin{figure*}
\includegraphics[angle=-90,width=16cm]{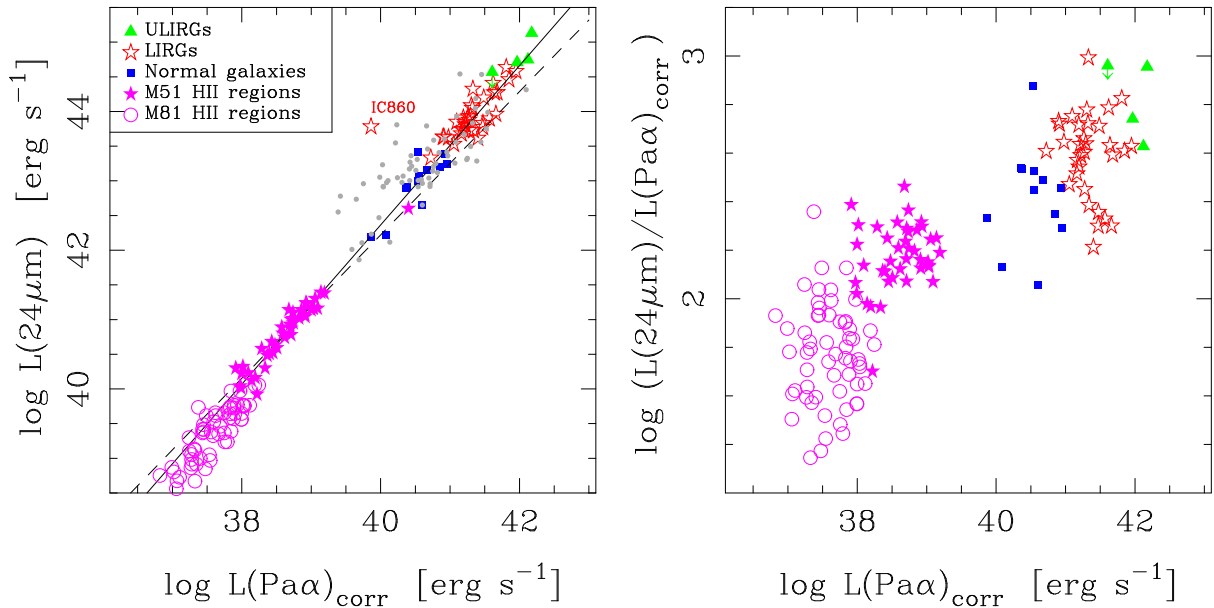}
\caption{{\it Left panel}. Monochromatic $24\,\mu$m
  luminosities vs. extinction-corrected Pa$\alpha$ luminosities 
for our sample of LIRGs, as well as normal
  galaxies and ULIRGs (those shown in Fig.~3). We also show in this comparison
  data for individual star-forming regions within two nearby galaxies with {\it
  Spitzer}/MIPS $24\,\mu$m
observations. For M51 the Calzetti et al. (2005) Pa$\alpha$ observations cover the
  central  6\,kpc of the galaxy. For M81 (see P\'erez-Gonz\'alez et al. 2006) the
  H$\alpha$ observations include the entire galaxy, and have been converted to
  Pa$\alpha$ luminosities assuming Case B recombination. 
The solid line is the derived fit to the M51
  knots, normal galaxies, LIRGs (excluding IC~860), and ULIRGs (see text). For
  comparison the dashed line shows the fit to the M51 H\,{\sc ii} knots given
  in Calzetti et al. (2005). Note how star-forming galaxies from the  UCM 
  and NFGS surveys  (grey dots, not included in the fit) also follow the trend.
{\it Right panel}. Ratio of the $24\,\mu$m to extinction-corrected
  Pa$\alpha$ luminosities vs. the extinction-corrected Pa$\alpha$
  luminosity. We only show galaxies with Pa$\alpha$ imaging data, that is, the
  UCM and NFGS galaxies are excluded. Symbols are as
  in left panel.}
\end{figure*}

As also noted by Calzetti et al. (2005) there are variations in the mid-IR to
Pa$\alpha$ ratios from galaxy to galaxy, clearly shown in Fig.~4 (right
panel). In particular, the 
$L(24\,\mu{\rm m})/L({\rm Pa}\alpha)_{\rm corr}$ ratios of the M81 H\,{\sc ii}
regions  appear to follow a different trend in terms of the Pa$\alpha$
luminosity than the rest of galaxies and the M51 H\,{\sc ii} regions.
(P\'erez-Gonz\'alez et al. 2006 attribute the large scatter 
of $L(24\,\mu{\rm m})/L({\rm Pa}\alpha)_{\rm corr}$ in M81 to uncertain extinction corrections).
The variation of 
the $L(24\,\mu{\rm m})/L({\rm Pa}\alpha)_{\rm corr}$ ratio
with the Pa$\alpha$ luminosity (i.e., SFR) 
between the M51 H\,{\sc ii} knots and our LIRGs and  ULIRGs is similar to the   
$L_{24\,\mu{\rm m}}$/SFR ratios found by Calzetti et al. (2005). 
In fact, the gap between the M51 H\,{\sc ii}
knots and the LIRGs/ULIRGs seen in our figure is occupied by their UV-selected
sample of starbursts, and the UCM and NFGS galaxies. 
We note however that we find a more pronounced variation of the
$L(24\,\mu{\rm m})/L({\rm Pa}\alpha)_{\rm corr}$ ratio because we use the
Pa$\alpha$ luminosity as a proxy for the SFR for LIRGs, whereas they use the
IR luminosity (see previous section, and Table~6). 

\subsubsection{Deeply embedded sources}

The deviation from strict proportionality in the $L(24\,\mu{\rm m})$ vs. 
$L({\rm Pa}\alpha)_{\rm corr}$ relation in dusty systems probably results from the
effects of extinction.
LIRGs and ULIRGs are known to contain highly obscured regions, 
usually associated with the nuclei of the galaxies,  that are
optically thick in the optical, and even in the near- and mid-IR (e.g, Genzel et
al. 1995; 1998; Alonso-Herrero et al. 
2000; Doyon et al. 1994; Kotilainen et al. 1996; Zhou et al. 1993). 
There is also indirect evidence based on the observed ratios of mid-IR
fine-structure emission lines that many of the youngest stars in massive
starbursts may still be
embedded in ultra-compact high density H\,{\sc ii} regions (Rigby \& Rieke
2004) and thus hidden from us by large amounts of extinction.

The departures from a constant $L(24\,\mu{\rm m})/L({\rm Pa}\alpha)_{\rm
  corr}$ ratio  may be linked to differences in the physical conditions in regions
  experiencing intense star formation. In particular, the average extinction for the
H\,{\sc ii} regions in M81 is $A_V \sim 0.8\,$mag (P\'erez-Gonz\'alez et al. 2006), whereas 
for the H\,{\sc ii} knots of M51 is $A_V \sim 3.5\,$mag (Calzetti et al. 2005), and for our
sample of LIRGs the average extinctions over the Pa$\alpha$ emitting regions are
$A_V \sim 2-6\,$mag. 

Since the estimated extinctions for the M51 H\,{\sc ii} regions and
  LIRGs are not too different, it is possible that we are underestimating the
  extinction in our sample of LIRGs. As shown for Arp~299
by Alonso-Herrero et al. (2000) and NGC~7469 by Genzel et al. (1995), the $J-H$
continuum colors tend to underestimate the extinction in very dusty systems. 
Thus, it is possible that the
photometric extinctions derived for some of the galaxies in our sample are
only lower limits to the true extinction (e.g., compare the photometric and
spectroscopic extinctions to the central regions of IRAS~17138$-$1017 and
NGC~2388). Particularly, this may be the case for the Pa$\alpha$ emission in
highly inclined dusty galaxies (those described in \S5.5).
We note also that not all of the most luminous IR systems in our 
sample are dominated by such extincted sources, for
instance, the star formation properties of the merger NGC~1614 can be 
explained with $A_K\simeq 0.5\,$mag (see Alonso-Herrero et al. 2001). 
Fortunately, the
  relative insensitivity of Pa$\alpha$ to extinction makes these differences
  relatively unimportant in estimating the SFR, or the scatter in Fig.~4 would
  be much larger than it is.

The high extinction indicated for the LIRGs, the possibility that the
extinction is actually higher than our estimates, and the likelihood that
some of the youngest stars are still embedded in ultracompact H\,{\sc ii} regions
are all consistent with our finding of a slope to the $L(24\mu{\rm m})$ vs.
$L({\rm Pa}\alpha)_{\rm corr}$ relation significantly larger than unity. We believe that
the physical explanation for this behavior is that the dust competes
increasingly effectively for ionizing photons and ultraviolet continuum
photons in very heavily obscured systems, so that there is a trend for an
increasing fraction of the luminosity to emerge in the IR. A
surprising result, however, is that this trend is so similar from galaxy
to galaxy. For reasonably dust embedded galaxies and H\,{\sc ii} regions, there is a linear
empirical relationship between IR and hydrogen recombination
luminosity with very small scatter that holds over more than four decades
in luminosity.

\subsubsection{Presence of AGN emission}
The presence of a bright AGN could make a significant contribution 
to the mid-IR luminosity of the system. Approximately 25\% of the galaxies are
classified as AGN  (Seyfert or  LINER) on the basis of optical line ratios (see Table~1). 
Three additional LIRGs (ESO~320-G030, IC~860, and Zw~049.057) present 
megamaser emission (Norris et al. 1986; Baan,
Haschick, \& Uglesich 1993). In extragalactic sources 
megamaser emission is generally
associated with a strong compact radio continuum source,  and about 70\% of megamaser
sources are classified as AGN or composite sources from optical spectroscopy (Baan et al. 1998), 
or tend to appear in galaxies with warmer {\it IRAS} colors (Baan et
al. 1993). Including the galaxies with megamaser emission the fraction of AGN in
our sample is consistent with that derived by other works (e.g., Veilleux et
al. 1995; Tran et al. 2001) for the same IR luminosity range.

For NGC~7469  Genzel et al. (1995) estimated that the
nucleus contributes up to 40\% of {\it IRAS} IR luminosity (see also Soifer
et al. 2003). For the B1 nucleus (that classified as a Seyfert) of NGC~3690
Keto et al. (1997) estimated a contribution of $20-30\%$ in the mid-IR. 
For the rest of the galaxies in our sample containing an AGN, only in the case
of IC~4518W we find that the Seyfert nucleus dominates 
the observed mid-IR emission (see Alonso-Herrero et al. 2006
in preparation).

Kewley et al. (2000) have argued, based on the
radio properties of a sample of warm IR galaxies with IR luminosities
similar to our sample, that in the cases where an
AGN is detected spectroscopically, it 
is rarely the dominant power source for the IR luminosity. 
In most cases in the galaxies in our sample containing an AGN there is extended H\,{\sc ii}
emission over large scales (e.g., NGC~7130; NGC~3690), so it is likely that the AGN contribution
to the total IR luminosity of the system is significantly smaller than in the
case of NGC~7469 and IC~4518W.

\subsubsection{The empirical calibration}
We can use the relation in equation (1) for empirical estimates of the
SFR in dusty environments. The most direct application is
to calculate the intrinsic extinction-corrected Pa$\alpha$ luminosity from
the monochromatic luminosity at $24\,\mu$m. Inverting equation (1), we find: 

\begin{equation}
L{\rm (Pa}\alpha)_{\rm corr} = 1244 \times L(24\mu{\rm m})^{0.871}.
\end{equation}

Kennicutt (1998) derived a relationship between H$\alpha$ luminosity and
the SFR. He also converted this expression to one between
total IR luminosity and SFR in dusty galaxies, and
this latter expression has been used widely to interpret  IR
observations. We will now update the expression based on the results
reported in this paper. First, we use monochromatic $24\,\mu$m
luminosities in place of total IR luminosities. Kennicutt (1998) assumed
that the great majority of the luminosity from young stars would be
absorbed by dust and reradiated in the far-IR in deriving his
relationship. Although this assumption is likely to be correct,
observationally there may be other contributions to the total IR
luminosity from older populations of stars or other luminosity sources
(e.g., review by Tuffs \& Popescu 2005). Therefore, a true total IR
luminosity measurement may overestimate the recent star formation in a
galaxy, although estimates based on {\it IRAS} measurements alone are less
subject to these issues because they only poorly sample the output of the
cold dust (because the longest band is at 100$\mu$m). Second, Kennicutt
assumed a direct proportionality between the H$\alpha$ luminosity and the
total IR. We find empirically that the increasing absorption
efficiency in increasingly luminous and obscured galaxies leads to a
deviation from strict proportionality toward increasing IR output
with increasing star-forming luminosity.

We begin by calibrating the mid-IR/Pa$\alpha$ relation using the H\,{\sc ii}
regions in M51. The extinction to these regions is roughly 2 magnitudes at
H$\alpha$. Since the morphologies are similar between H$\alpha$ and
Pa$\alpha$ (Quillen \& Yukita 2001), a
straightforward extinction correction is appropriate (e.g., there are
minimal issues with very heavily obscured regions that contribute no
H$\alpha$).  Assuming Case B recombination, 
the relation between SFR and H$\alpha$ quoted by Kennicutt (1998)
can be converted to 
SFR(M$_\odot$ yr$^{-1}$) = $6.79\times 10^{-41} \, L({\rm Pa}\alpha)$
(erg s$^{-1}$). From these considerations, we derive the following  relation between the SFR rate and mid-IR
luminosity for luminous, dusty galaxies is:

\begin{equation}
{\rm SFR}({\rm M}_\odot\,{\rm yr}^{-1}) = 8.45 \times 10^{-38}\, (L(24\,\mu{\rm m})/
({\rm erg\,s}^{-1}))^{0.871}.    
\end{equation}

\noindent This relation is analogous to the widely used relation between SFR and IR
luminosity (Kennicutt 1998) but it is not affected by the uncertain
contribution to the total IR luminosity of a galaxy of 
dust heating from old
stars.

\section{Summary}
We have analyzed {\it HST}/NICMOS $1.1-1.89\,\mu$m 
continuum and Pa$\alpha$ emission line observations
of a volume limited sample of 30 local universe LIRGs ($\log L_{\rm
  IR}=11-11.9\,[{\rm L}_\odot]$).  The galaxies have been selected so that the Pa$\alpha$ emission
line could be observed with the NICMOS F190N narrow-band filter ($2800 <v <
5200\,$km s$^{-1}$). The sample 
comprises approximately 80\% of all the LIRGs in the RBGS
 in the above velocity interval, and is representative
of local LIRGs in general. 
The NICMOS observations cover the central $\simeq 3.3-7.2\,$kpc of the galaxies.
We find the following:

\begin{enumerate}

\item
The most common morphological continuum features in 
the central regions are bright star clusters, and large-scale 
spiral arms, sometimes extending  
down to the inner kpc (mini-spirals). 
In highly inclined systems there are dust lanes crossing the disks of the
galaxies and in some cases, even hiding the nucleus of the galaxy.
The  ``extreme'' (perturbed) morphologies commonly seen in ULIRGs are only
observed in some of the most IR luminous examples in our sample. 

\item
Approximately half of the LIRGs show compact ($\sim
1-2\,$kpc) Pa$\alpha$ morphologies in the form of nuclear star formation
rings, mini-spiral structure, and emission
associated with the nucleus of the galaxy. The typical observed (not corrected
for extinction) H$\alpha$
surface
brightnesses are between $2-10 \times 10^{41}\,{\rm erg \, s}^
{-1}$ kpc$^{-2}$, although the most extreme cases, 
for instance the ring of star formation
of NGC~1614, can reach $\simeq 60 \times 10^{41}\,{\rm erg \, s}^
{-1}$ kpc$^{-2}$.

\item
The remaining half of the galaxies 
show Pa$\alpha$ emission extending over scales of $\simeq 3.3-7.2\,$kpc and
larger, with bright H\,{\sc ii} regions in the spiral arms (face-on systems) or
along the disks of the galaxies (edge-on systems), with or without
bright nuclear emission. The H$\alpha$ surface brightnesses are about one
order of magnitude fainter than in galaxies with compact Pa$\alpha$ emission.

\item
Most of the LIRGs show a population of numerous bright H\,{\sc ii}
regions, and in about half of them the typical H\,{\sc ii} region 
(the median H$\alpha$ luminosity) is as bright as or
brighter than the
giant H\,{\sc ii} region 30 Dor. The most IR  (as well as in H$\alpha$) luminous
galaxies tend to host the brightest median and  
first-ranked H\,{\sc ii} regions.

\item
The extinctions to the gas (derived from the 
H$\alpha$/Pa$\alpha$ and Pa$\alpha$/Br$\gamma$ line ratios) and 
to the stars (derived from the  
$m_{\rm F110W}-m_{\rm F160W}$ colors) are 
on average 
$A_V\sim 2-6\,$mag over the Pa$\alpha$ emitting regions covered by the {\it
  HST}/NICMOS images.

\item
There exists a good correlation between the mid-IR $24\,\mu$m and the
extinction-corrected Pa$\alpha$ luminosities of LIRGs, ULIRGs, normal
galaxies, and H\,{\sc ii} regions of M51, covering nearly five decades in
luminosity.  This suggests that the mid-IR
luminosity of galaxies undergoing dusty, intense star formation is a good indicator
of the SFR. The increasing 
$L(24\,\mu{\rm m})/L({\rm Pa}\alpha)_{\rm corr}$ ratio for more luminous
Pa$\alpha$ emitters (that is, galaxies with  higher SFRs) 
may be due to the presence of deeply
embedded sources in the youngest 
star-forming regions for which we have underestimated the extinction, 
even at near-IR wavelengths.

\item 
Analogous to the widely used relation between SFR and total
IR luminosity (Kennicutt 1998), we show that  SFRs
can be determined accurately in luminous, dusty galaxies as:

SFR(M$_\odot$ yr$^{-1}) = 8.45 \times 10^{-38}$
(L(24$\mu$m)/(erg s$^{-1}$))$^{0.871}$

\end{enumerate}

$\,$\\

Based on observations with the NASA/ESA Hubble Space
Telescope at the Space Telescope Science
Institute, which is operated by the Association of Universities for
Research in Astronomy, Inc., under NASA contract NAS 5-26555. 
This research has made use of the NASA/IPAC Extragalactic Database (NED),
which is operated by the Jet Propulsion Laboratory, California Institute of
Technology, under contract with the National Aeronautics and Space
Administration (NASA). 

We thank an anonymous referee for a careful reading of the manuscript and useful
suggestions. 
We would like to thank D. Calzetti and R. Kennicutt for enlightening
discussions. 
We are grateful to T. Hattori for providing us with
 H$\alpha$ images of the galaxies. We thank T. D\'{\i}az-Santos for helping us with the {\it
  Spitzer}/MIPS images of the sample of LIRGs.
AAH and LC acknowledge support from 
the Spanish Programa Nacional de Astronom\'{\i}a y Astrof\'{\i}sica 
under grant AYA2002-01055 and Plan Nacional del Espacio ESP2005-01480, 
and PGPG from the Spanish Programa Nacional
de Astronom\'{\i}a y Astrof\'{\i}sica under grant AYA 2004-01676. This work has been funded by NASA grant
HST-GO-10169 and by NASA through contract 1255094 issued by JPL/Caltech.


\begin{thebibliography}{101}

\bibitem{b1}
Alonso-Herrero, A., Rieke, G. H., Rieke, M. J., \&
Scoville N. Z. 2000, ApJ, 532, 845 

\bibitem{b2}Alonso-Herrero, A., Engelbracht, C. W., 
Rieke, M. J., Rieke, G. H., \& Quillen, A. C. 2001, ApJ, 546,  952 

\bibitem{b3}Alonso-Herrero, A., \& Knapen, J. H. 2001, AJ, 122, 1350

\bibitem{b4}Alonso-Herrero, A., Rieke, G. H., Rieke, M. J., 
\& Scoville, N. Z. 2002, AJ, 124, 166

\bibitem{b5} Arribas, S., Bushouse, H., Lucas, R. A., Colina, L, \& Borne,
  K. D.  et al. 2004, AJ, 127, 2522


\bibitem{b7}Baan, W. A., Haschick, A. D., \& Uglesich, R. 1993, ApJ, 415, 140

\bibitem{b8}Baan, W. A., Salzer, J. J., \& LeWinter, R. D. 1998, ApJ, 509, 633

\bibitem{b9}Beckman, J. E., Rozas, M., Zurita, A., Watson, R. A., \& Knapen,
  J. H. 2000, AJ, 119, 2728

\bibitem{b10}Bell, E. F. et al. 2005, ApJ, 625, 23

\bibitem{b11}
Bertin, E. \& Arnouts, S. 1996, A\&AS, 117, 393

\bibitem{b12}Bland-Hawthorn, J., \& Maloney, P. R. 1999, ApJ, 510, L33 

\bibitem{b13} B\"oker, T. et al. 1999, ApJS, 124, 95

\bibitem{b14} B\"oker, T., Sarzi, M., McLaughlin, D. E., van der Marel, R.,
  Rix, H.-W., Ho, L. C., \& Shields, J. C. 2004, AJ, 127, 105

\bibitem{b15} Bushouse, H. A., Borne, K. D., Colina, L., Lucas, R. A.,
  Rowan-Robinson, M., Baker, A. C., Clements, D. L., Lawrence, A., \& Oliver,
  S. 2002, ApJS, 138, 1

\bibitem{b16}Calzetti, D., Kinney, A, L., \& Storchi-Bergmann, T. 1994, 
ApJ, 429, 582 

\bibitem{b17}Calzetti, D. et al. 2005, ApJ, 633, 871

\bibitem{b18}Clements, D. L., Sutherland, W. J., McMahon, R. G., 
\& Saunders, W. 1996, MNRAS, 279, 477

\bibitem{b19}Colina, L., Arribas, S., \& Clements, D. 2004, ApJ, 602, 181


\bibitem{b20}Colina, L., Borne, K., Bushouse, H., Lucas, R. A.,
  Rowan-Robinson, M., Lawrence, A., Clements, D., Baker, A., \& Oliver, S. 
2001, ApJ, 563, 546

\bibitem{b21}Corbett, E. A., Kewley, L., Appleton, P. N., Charmandaris, V.,
  Dopita, M. A., Heisler, C. A., Norris, R. P., Zezas, A., \& Marston, A. 
2003, ApJ, 583, 670

\bibitem{b22}Dale, D. A., \& Helou, G. 2002, ApJ, 576, 159


\bibitem{b24}Dopita, M. A., Pereira, M., Kewley, L. J., 
\& Capaccioli, M. 2002, ApJS, 143, 47

\bibitem{b25}Doyon, R., Joseph, R. D., \& Wright, G. S. 1994, ApJ, 421, 101

\bibitem{b26}Elbaz, D., Cesarsky, C. J., Chanial, P., Aussel, H.,
  Franceschini, A., Fadda, D., \& Chary, R. R. 2002, A\&A, 384, 848



\bibitem{b28}Garc\'{\i}a-Mar\'{\i}n, M., Colina, L, Arribas, S.,
  Alonso-Herrero, A., \& Mediavilla, E. 2006, ApJ, submitted

\bibitem{b29}Genzel, R., Weitzel, L., Tacconi-Garman, L. E., Blietz, M.,
  Cameron, M., Krabbe, A., Lutz, D., \& Sternberg, A. 1995, ApJ, 444, 129

\bibitem{b30}Genzel, R. et al. 1998, ApJ, 498, 579

\bibitem{b31}Genzel, R. et al. 2001, ApJ, 563, 527

\bibitem{b32}Gehrz, R. D., Sramek, R. A., \& Weedman, D. W.
1983, ApJ, 267, 551

\bibitem{b33}Goldader, J. D., Joseph, R. D., Doyon, R., 
\& Sanders, D. B. 1997a, ApJS, 108, 449

\bibitem{b34}Goldader, J. D., Joseph, R. D., Doyon, R., 
\& Sanders, D. B. 1997b, ApJ, 474, 104

\bibitem{b35}Hattori, T., Yoshida, M., Ohtani, H., Sugai, H., Ishigaki, T.,
  Sasaki, M., Hayashi, T., Ozaki, S., Ishii, M., \& Kawai, A. 2004, 
AJ, 127, 736


\bibitem{b36}Hummer, D. G., \& Storey, P. J. 1987, MNRAS, 224, 801



\bibitem{b37}Kennicutt, R. C. Jr., Edgar, B. K., \&
Hodge, P. W. 1989, ApJ, 337, 761

\bibitem{b38}Kennicutt, R. C. Jr. 1998, ARA\&A, 36, 189

\bibitem{b39}Keto, E. et al. 1997, ApJ, 485, 598

\bibitem{b40}Kewley, L. J., Heisler, C. A., Dopita, M. A., Sutherland, R.,
  Norris, R. P., Reynolds, J., \& Lumsden, S.  2000, ApJ, 530, 704

\bibitem{b41}Kewley, L. J., Heisler, C. A., Dopita, M. A., 
\& Lumsden, S. 2001, ApJS, 132, 37

\bibitem{b42}Kewley, L. J., Geller, M. J., Jansen, R. A., \& Dopita,
  M. A. 2002, AJ, 124, 3135

\bibitem{b43}Kim, D.-C., Sanders, D.B., Veilleux, S., Mazzarella, J.M., \&
  Soifer, B. T. 1995, ApJS, 98, 129 

\bibitem{b44}Kotilainen, J. K., Moorwood, A. F. M., Ward, M. J., 
\& Forbes, D. A. 1996, A\&A, 305, 107 

\bibitem{b45}Lagache, G., Puget, J.-L., \& Dole, H. 2005, ARA\&A, 43, 727 

\bibitem{b46}Le Floc'h, E. et al. 2005, ApJ, 632, 169

\bibitem{b47}Lehnert, M. D. \& Heckman, T. M. 1995, ApJS, 97, 89
\bibitem{b48}Leitherer, C. et al. 1999, ApJS, 123, 3

\bibitem{b49}L\'{\i}pari, S., D\'{\i}az, R., Taniguchi, Y., Terlevich, R.,
  Dottori, H., \& Carranza, G. 2000, AJ, 120, 645

\bibitem{b50}L\'{\i}pari, S., D\'{\i}az, R. J., Forte, J. C., Terlevich, R.,
  Taniguchi, Y., Aguero, M. P., Alonso-Herrero, A., Mediavilla, E., 
\& Zepf, S. 2004, MNRAS, 354, L1 

\bibitem{b51}McLeod, K. K., Rieke, G. H., Rieke, M. J., \& Kelly, D. M. 1993,
  ApJ, 412, 111

\bibitem{b52}Maoz, D., Barth, A. J., Ho, L. C., Sternberg, A., \& Filippenko,
  A. V. 2001, AJ, 121, 3048

\bibitem{b53}M\'arquez, I., Moles, M.,  \& Masegosa, J. 1996, A\&A, 310, 401

\bibitem{b54}Martini, P., Regan, M. W., Mulchaey, J. S., \& 
Pogge, B. W. 2003, ApJ, 589, 774

\bibitem{b55}Melbourne, J. \& Salzer, J. J. 2002, AJ, 123, 2302

\bibitem{b56}Melbourne, J., Koo, D. C., \& Le Floc'h, E. 2005, ApJ, 632, 65

\bibitem{b57}Miles, J. W., Houck, J. R., Hayward, T. L., \& Ashby,
  M. L. N. 1996, ApJ, 465, 191

\bibitem{b58}Murphy, T. W., Soifer, B. T., Matthews, K., 
\& Armus, L. 2001, ApJ, 559, 201

\bibitem{b59}Murphy, T. W. Jr., Armus, L.,  Matthews, K., Soifer, B. T.,
  Mazzarella, J. M., Shupe, D. L., Strauss, M. A.  
\& Neugebauer, G. 1996, AJ, 111, 1025


\bibitem{b61}Norris, R. P., Whiteoak, J. B., Gadner, F. F., Allen, D. A., 
\& Roche, P. F. 1986, MNRAS, 221, 51P

\bibitem{b62}
Papovich, C., Egami, E., Le Floc'h, E., P\'erez-Gonz\'alez, P., Rieke, G.,
Rigby, J., Dole, H., \& Rieke, M. 2005, in press,   proceedings of the 
STScI May 2004 Symposium, "Planets to Cosmology: 
Essential Science in Hubble's Final Years"
(astro-ph/0408454)

\bibitem{b63}P\'erez-Gonz\'alez, P. G., Zamorano, J., Gallego, J.,
  Arag\'on-Salamanca, A., \& Gil de Paz, A. 2003, ApJ, 591, 827

\bibitem{b64}P\'erez-Gonz\'alez, P. G. et al. 2005, ApJ, 630, 82

\bibitem{b65}P\'erez-Gonz\'alez, P. G. et al. 2006, submitted

\bibitem{b66}Quillen, A. C., McDonald, C., Alonso-Herrero, A., Lee, A.,
  Shaked, S., Rieke, M. J., \& Rieke, G. H. 2001, ApJ, 547, 129 

\bibitem{b67}Quillen, A. C., \& Yukita, M. 2001, AJ, 121, 2095



\bibitem{b69}Rieke, G. H. \& Low, F. J. 1972, ApJ, 176, L95

\bibitem{b70}Rieke, G. H. \& Lebofsky, M. J. 1985, ApJ, 288, 618

\bibitem{b71}Rieke, G. H., Loken, K., Rieke, M. J., 
\& Tamblyn, P. 1993, ApJ, 412, 99

\bibitem{b72}Rigby, J. R., \& Rieke, G. H. 2004, ApJ, 606, 237

\bibitem{b73}Roy, A. L., Goss, W. M., Mohan, N. R.,
\& Anantharamaiah, K. R. 2005, A\&A, 435, 831

\bibitem{b74}Rush, B., Malkan, M. A., \& Spinoglio, L. 1993,
ApJS, 89, 1

\bibitem{b75}Sanders, D. B. \& Mirabel, I. F. 1996, ARA\&A, 34, 749

\bibitem{b76}Sanders, D. B., Soifer, B. T., Elias, J. H., 
Neugebauer, B., \& Matthews, K. 1988, ApJ, 328, L35

\bibitem{b77} Sanders, D. B., Egami, E., L\'{\i}pari, S., 
Mirabel, I. F., \& Soifer, B. T. 1995, AJ, 110, 1993

\bibitem{b78}Sanders, D. B., Mazzarella, J. M., Kim, D.-C., 
Surace, J. A., \& Soifer, B. T. 2003, AJ, 126, 1607

\bibitem{b79}Satyapal, S., Watson, D. M., Pipher, J. L., Forrest, W. J., 
Fischer, J., Greenhouse, M. A., Smith, H. A., \& Woodward, C. E. 
1999, ApJ, 516, 704

\bibitem{b80}Scoville, N. Z. et al. 2000, AJ, 119, 991

\bibitem{b81}Scoville, N. Z., Poletta, M., Ewald, S., Stolovy, S. R.,
  Thompson, R., \& Rieke, M. J. 2001, AJ, 122, 3017

\bibitem{b82}Sekiguchi, K. \& Wolstencroft, R. D. 1992, A\&A, 255, 581


\bibitem{b83}Shi, Y., Rieke, G. H., Papovich, C., P\'erez-Gonz\'alez, P. G.,
  \& Le Floc'h, E. 2006, ApJ, in press (astro-ph/0603453)
\bibitem{b84}Soifer, B. T., Sanders, D. B., Madore, B. F., Neugebauer, G.,
  Danielson, G. E., Elias, J. H., Lonsdale, C. J., \& Rice, W. L.s 1987, ApJ, 320, 238

\bibitem{b85}Soifer, B. T., Neugebauer, G. 
Matthews, K., Egami, E., Weinberger, A. J.,  Ressler, M., 
Scoville, N. Z., Stolovy, S. R., Condon, J. J., 
\& Becklin, E. E. 2001, AJ, 122, 1213

\bibitem{b86}Soifer, B. T., Bock, J. J., Marsh, K., 
Neugebauer, G., Matthews, K., Egami, E., \& Armus, L. 
2003, AJ, 123, 146  

\bibitem{b87}Sugai, H., Davies, R. I., Malkan, M. A., McLean, I. S., 
Usuda, T., \& Ward, M. J. 1999, ApJ, 527, 778

\bibitem{b88}Surace, J. A., Sanders, D. B., \& Mazzarella, J. M. 2004, AJ,
  127, 3235

\bibitem{b89}Surace, J. A., Sanders, D. B., \& Evans, A. S. 2000, ApJ, 529,
  170 

\bibitem{b90}Takeuchi, T. T., Buat, V., Iglesias-P\'aramo, J., 
Boselli, A., \& Burgarella, D. 2005, A\&A, 432, 423

\bibitem{b91}Tran, Q. D. et al. 2001, ApJ, 552, 527

\bibitem{b92}Tuffs, R. J., \& Popescu, C. c. 2005, AIP Conf. Ser. 216, 84


\bibitem{b93}van den Broek, A. C. et al. 1991, A\&AS, 91, 61

\bibitem{b94}Veilleux, S., Kim, D.-C., Sanders, D. B., Mazzarella, J. M., \& Soifer,
B. T. 1995, ApJS, 98, 171


\bibitem{b95}V\'eron-Cetty, M.-P. \& V\'eron, P. 2001, A\&A, 374, 92

\bibitem{b96}Witt, A. N. \& Gordon, K. D. 2000, ApJ, 528, 799

\bibitem{b97}Wu, H., Zou, Z. L., Xia, X. Y., \& Deng, Z. G. 1998, A\&AS, 132, 181

\bibitem{b98}Wynn-Wiliams, C. G., \& Becklin, E. E. 
1993, ApJ, 412, 535

\bibitem{b99}Zhao, J.-H., Anantharamaiah, K. R., Goss, W. M., 
\& Viallefond, F. 1997, ApJ, 482, 186

\bibitem{b100}Zheng, X. Z., Hammer, F., Flores, H., Ass\'emat, F., \& Pelat,
  D. 2004, A\&A, 421, 847

\bibitem{b101}Zhou, S., Wynn-Williams, C. G. \& Sanders, D. B. 1993, ApJ, 409,
  149 

\end{thebibliography}
\end{document}